\documentclass[11pt,a4paper]{article}      % Specifies the document class
\pdfoutput=1

\usepackage{amsthm,amsmath,amssymb,xcolor,sectsty,latexsym,mathtools,hyperref}
\usepackage{slashed}  
\usepackage{bm}
\usepackage{epsfig}
\usepackage{dcolumn}
\definecolor{vermelho}{cmyk}{0,.88,.77,.40}
\hypersetup{urlcolor=black, colorlinks=true,linkcolor=vermelho}
\usepackage{cite}
\usepackage{feynmp}
\numberwithin{equation}{section}
\usepackage[textwidth=6.3in,top=1.3in,bottom=1.3in]{geometry}
\usepackage{setspace}
\chapterfont{\color{vermelho}}
\usepackage{simplewick}
\usepackage{hyperref}
\hypersetup{
    colorlinks   = true,
    citecolor    = blue
}

\usepackage[utf8]{inputenc}
\newcommand{\be}{\begin{equation}}
\newcommand{\ee}{\end{equation}}
\newcommand{\beq}{\begin{equation}}
\newcommand{\eeq}{\end{equation}}
\newcommand{\ba}{\begin{eqnarray}}
\newcommand{\ea}{\end{eqnarray}}
\newcommand{\bef}{\begin{figure}}
	\newcommand{\eef}{\end{figure}}

\newcommand{\p}{\partial}

\newcommand{\e}{\epsilon}

\newcommand{\g}{\gamma}

\newcommand{\cO}{{\cal O}}

\newcommand{\cL}{{\cal L}}

\newcommand{\cI}{{\cal I}^+}

\def\x{{\bf x}}

\newcommand{\nn}{\nonumber}
\newcommand{\nc}{\text{NC}}
\newcommand{\dg}{\dot{\gamma_k}}

\begin{document}

\thispagestyle{empty}
\begin{titlepage}
\nopagebreak

\title{  \begin{center}\bf Asymptotic Symmetries in\\ de Sitter and Inflationary Spacetimes \end{center} }

\vfill
\author{Ricardo Z. Ferreira$^{a,b}$\footnote{ferreira@cp3.sdu.dk}, ~ McCullen Sandora$^{a,c}$\footnote{sandora@cp3.sdu.dk}~ and ~  Martin S. Sloth$^{a}$\footnote{sloth@cp3.sdu.dk}
}
\date{ }

\maketitle

\begin{center}
	\vspace{-0.7cm}
	{\it  $^a$CP$^3$-Origins, Center for Cosmology and Particle Physics Phenomenology}\\
	{\it  University of Southern Denmark, Campusvej 55, 5230 Odense M, Denmark}\\
	\vspace{0.2cm}
	{\it  $^b$Departament de Fisica Fonamental i Institut de Ciencies del Cosmos}\\
	{\it  Universitat de Barcelona, Marti i Franques, 1, 08028, Barcelona, Spain}\\
	\vspace{0.2cm}
	{\it  $^c$Institute of Cosmology, Department of Physics and Astronomy}\\
	{\it  Tufts University, Medford, MA 02155, USA}
	
\end{center}
\vfill
\begin{abstract}
Soft gravitons produced by the expansion of de Sitter can be viewed as the Nambu-Goldstone bosons of spontaneously broken asymptotic symmetries of the de Sitter spacetime. We explicitly construct the associated charges, and show that acting with the charges on the vacuum creates a new state equivalent to a change in the local coordinates induced by the soft graviton.  While the effect remains unobservable within the domain of a single observer where the symmetry is unbroken, this change is physical when comparing different asymptotic observers, or between a transformed and un-transformed initial state, consistent with the scale-dependent statistical anisotropies previously derived using semiclassical relations. We then compute the overlap, $\langle0| 0'\rangle$, between the unperturbed de Sitter vacuum $|0\rangle$, and the state $| 0'\rangle$ obtained by acting $\mathcal{N}$ times with the charge. We show that when $\mathcal{N}\to M_p^2/H^2$ this overlap receives order one corrections and $\langle0| 0'\rangle\to 0$, which corresponds to an infrared perturbative breakdown after a time $t_{dS} \sim M_p^2/H^3$ has elapsed, consistent with earlier arguments in the literature arguing for a perturbative breakdown on this timescale. We also discuss the generalization to inflation, and rederive the 3-point and one-loop consistency relations. 
\end{abstract}
\noindent
DNRF90
\hfill \\
\vfill
\end{titlepage}

\section{Introduction}
Three of the most important problems in fundamental cosmology, the quantum origin of the universe, the cosmological constant problem, and the black hole information paradox, are all believed to have to do with infrared problems in quantum gravity. Assuming that there cannot be three unrelated infrared problems of quantum gravity indicates that they must somehow be related. Thus, in addition to trying to tackle these problems individually, it makes sense to also explore their interconnection, in the hope that identifying their commonalities will shed light on what their resolutions might be.

A possible connection between the cosmological constant problem and the quantum origin of the universe has been known for some time. In the landscape, the smallness of the observed cosmological constant appears as a consequence of environmental selection \cite{Weinberg:1988cp}, and eternal inflation is commonly invoked as a means of populating all false vacua and thereby realizing all possible universes\cite{hep-th/0302219, hep-th/0105097,Bousso:2000xa,Linde:1983gd}.  

A concrete connection between black holes and eternal inflation is so far missing, although there are many similarities between Schwarzschild and de Sitter spacetimes.  On a heuristic level, they both have horizons, and there are indications of a perturbative breakdown due to the presence of infrared modes in both spacetimes\footnote{It has been suggested  that this could also provide a different connection between inflation and the cosmological constant problem \cite{na-h,Polyakov:1982ug, 0709.2899, 0912.5503, Tsamis:1992sx, Tsamis:1994ca,0708.2004}, although this is not the focus of the present work.} on parametrically similar timescales \cite{hep-th/0703116, 0911.3395, 1005.1056, 1104.0002, 1109.1000, Dvali:2013eja,Dvali:2014gua}. Below we review some of the similarities between the two situations, and then in the main part of the paper, we elaborate on the connection by reanalyzing the infrared issues in de Sitter from the point of view of asymptotic symmetries, which have recently been argued to be important for the understanding of the black hole evaporation process.

\subsection{Black Hole Evaporation}

In order to make the connection between the dynamics of black hole evaporation and the infrared issues in de Sitter space as tight as possible, let us first review the black hole information paradox very briefly with a focus on the connection to the cosmological context. Perturbatively, one can view Hawking radiation as pairs of quantum particles created at the horizon, where one escapes to infinity and the other falls into the black hole. The escaping quantum particle is entangled with the interior quantum particle, and its entanglement entropy is
\beq
S_{ent} = \ln 2~.
\eeq
In the course of its evolution, the black hole emits $\mathcal{N}_{bh}\sim (M/M_{p})^2$ such quanta before it reaches a size of order $l_{p}$, where $M$ is the mass of the black hole, and $M_p$ and $l_p$ are the Planck mass and length respectively.  At the end of its evolution it will then have an entanglement entropy
\beq
S_{ent} \gtrsim \mathcal{N}_{bh} \ln 2~.
\eeq
At this point, if the Planck sized black hole evaporates away, the external emitted particles will make up the full quantum system with an entropy given by $S_{ent}$, indicating that they are in a mixed state, despite the fact that they might have been created initially in a pure state, violating unitarity. On the other hand, a Planck sized remnant with arbitrarily high degeneracy would be a strange quantum state, with unacceptable physical consequences \cite{Giddings:1994qt,Susskind:1995da}. 

Assuming that both of these possibilities are unacceptable, it has been argued that the apparent paradox is a consequence of trusting perturbation theory beyond its limits of validity. Different scenarios for a non-perturbative description of black hole evaporation have been put forward (some recent examples are \cite{Mathur:2003hj,1108.2015,Giddings:2012gc,1306.0533}). In order to shed light on the problem from a different angle, it might be useful to understand the exact source of the perturbative breakdown within perturbation theory. Since we know that the black hole entropy is proportional to the horizon area, and given the symmetry between the entanglement entropy of interior versus exterior modes, it is clear that the growth in entanglement entropy has to halt when it reaches the size of the total black hole entropy and becomes decreasing as the black hole starts to evaporate, on the timescale given by the black hole half-life. At this point, information needs to start coming out of the black hole, as first argued by Page \cite{hep-th/9306083}, and the simple perturbative arguments above are therefore expected to break down. The entanglement entropy computed by the simple perturbative argument above exceeds the black hole entropy when the amount of emitted quanta is
\beq
\mathcal{N}_{bh} \sim S_{BH}/\ln 2 = (M/M_{p})^2/(2\ln 2)~, 
\eeq
which is the amount of quanta emitted on the timescale of the black hole evaporation time. Using similar reasoning, it was argued in \cite{na-h} and \cite{hep-th/0703116} that we should expect a breakdown in the perturbative description of black hole evaporation on this time scale,
\beq \label{tev}
t_{ev} \sim R_{S} S_{BH} \sim M^3/M_p^4,
\eeq 
when order $\mathcal{N}_{bh}$ quanta have been emitted.

How would one diagnose such a perturbative breakdown? In perturbation theory, the evolution of the time-dependent perturbed quantum state of the black hole, $\left|\psi(t)\right>$ in the interacting picture, has the form
\beq
\left|\psi(t)\right> = T \exp\left[-i \int^t H_I(t)\right]\left|\psi\right>_0~,
\eeq
where $H_I$ is the interacting Hamiltonian, and $\left|\psi(t)\right>_0$ is the unperturbed state. Perturbation theory remains valid as long as the matrix elements of $\int^t H_I(t)$ remain small. It was therefore argued by Giddings in \cite{hep-th/0703116}, that if in a nice-slicing of the black hole background (where curvature is everywhere small on the spatial slices), we have 
\beq\label{olbh}
\langle\psi(t)|\psi\rangle_0\rightarrow0
\eeq
for some time $t$, then it would be an indicator that perturbation theory has broken down, and we might therefore expect this to happen on the time scale of $t_{ev}$.

%In the black hole context, the information paradox signals a perturbative breakdown on the Page time (REF), the time scale it information to come out of the black hole (REF). This breakdown transmutes the information paradox into merely an information problem, which is to understand the nonperturbative evolution process of black hole evaporation.  Recently, it has been proposed that this problem can be understood by considering soft hair \cite{1401.7026,1411.5745}. The addition of a zero-mode Hawking quantum (a soft mode), corresponds to a symmetry transformation of the metric at infinity in the form of  a large gauge transformation, or a supertranslation. It was argued that this does not leave vacuum state invariant, and that this change in the vacuum can encode the missing information. This was shown by constructing a conserved charge out of the large gauge transformation, and demonstrating that it acts nontrivially on the final state of the evaporation process.

\subsection{Quantum Gravity in de Sitter}

In de Sitter space information is not necessarily permanently lost when it recedes behind an observer's horizon.  If the cosmological constant decays, as happens at the end of inflation, information will in fact reenter the horizon, and thus there is no reason to believe that the Gibbons-Hawking radiation in de Sitter space will carry information.  Therefore, there is no exact analogue of the black hole information paradox in de Sitter, although the two space-times do have many similarities. An important point is that we also can associate entropy with the de Sitter horizon
\beq
S_{dS} = \frac{1}{4}\frac{A_{dS}}{G}= \frac{1}{8} R_{ds}^2 M_p^2~,
\eeq
where $R_{dS}$ is the de Sitter radius. It has been argued that this implies that the number of fundamental degrees of freedom in de Sitter, $N$,  is finite \cite{hep-th/0007146} and given by $S_{dS} = \ln N$. As noted in \cite{hep-th/0106109}, this assumption clearly implies that perturbation theory must break down in de Sitter in the IR, since an infinite de Sitter space has an infinite Hilbert space in perturbation theory. 

During every e-fold of de Sitter expansion, $\Delta t = 1/H$, two soft (super-horizon) gravitons are emitted at the de Sitter horizon. That means that after $N_e= \ln a(t)= Ht$ e-folds of expansion, roughly $\mathcal{N}_{dS} \sim N_e$ independent quanta are emitted making up $2^{\mathcal{N}_{dS}}$ possible microstates, which exceeds $N$ when
\beq
\mathcal{N}_{dS}  \sim S_{dS}/\ln(2)~.
\eeq
In fact it was argued in \cite{hep-th/0306070}, that the entropy of the emitted modes, $S_e$, can be viewed as the entanglement entropy of de Sitter, and it was further shown\footnote{See appendix A of \cite{hep-th/0306070}.} that the change in entropy per e-fold satisfies $dS_e/dN_e \gtrsim 1$, which when integrated gives \cite{hep-th/0306070,hep-th/0703116,ArkaniHamed:2007ky}, 
\beq
S_{e} \gtrsim N_e ~.
\eeq
From this point of view it seems reasonable to expect that perturbation theory will break down when $S_{e} \gtrsim S_{dS}$, which happens on a time scale dimensionally equivalent to the black hole evaporation time in de Sitter
\beq\label{tds}
t_{dS} \sim R_{dS} S_{dS} \sim H^{-3}M_p^2\,.
\eeq

It is interesting that it has also been argued how a finite Hilbert space could emerge in de Sitter. If we consider the overlap between initial states and final states, $\langle i | f \rangle$, in de Sitter as a kind of ``meta observable" (similar to the ``q-observable" discussed in \cite{1104.0002}), then the matrix defined by $\langle i | f \rangle$ has infinite dimension. But it  is argued that it can have finite rank if we mod out by all initial states $\langle i |$ for which $\langle i |f\rangle=0$ \cite{hep-th/0106109} (for instance by simply setting them to 0).  

Combining the thoughts above regarding a possible perturbative breakdown in de Sitter due to the finiteness of the Hilbert space, and the idea that finiteness shows up as $\langle i |f\rangle\to0$, it is interesting to compute an overlap like $\langle i | f \rangle$ in a finite part of de Sitter with a finite number of degrees of freedom and then see that $\langle i | f \rangle \to 0$ as we remove the IR cutoff and include more and more degrees of freedom. We then expect that once $\mathcal{N}_{dS} \sim M_p^2/H^2$ quanta has been emitted, perturbation theory will break down as a consequence of
\beq \label{olds}
\langle i | f \rangle \to 0
\eeq
 at this point. Note that this is happens naturally after the time $t_{dS}$, showing the equivalence of these two different diagnostics for the perturbative breakdown.

 \subsection{Diagnosing the Perturbative IR Breakdown}

Given the striking similarities between the black hole information paradox and the entropy problem of de Sitter in (\ref{tev}) and (\ref{tds}), one might speculate that they are fundamentally two sides of the same problem, and we might try to diagnose the exact source of the perturbative breakdown by looking at the non-linear evolution of perturbations on timescales given by $t\sim R S$. De Sitter space enjoys many more symmetries than black holes, and therefore progress is easier here.  This was done in a series of papers \cite{1005.1056,1104.0002,1109.1000}. In \cite{1005.1056}, a perturbative breakdown in cosmological correlation functions was identified on this exact timescale.  In \cite{1104.0002} the physical meaning of such a perturbative breakdown was examined more carefully, and the relation between local observers and global observers was discussed. Finally in \cite{1109.1000}, the authors used the fluctuations of geodesics as a gauge invariant measure of the geometry in de Sitter, and showed that these become large, signalling breakdown of a perturbative description of the geometry, consistent with perturbative instability of de Sitter space. In fact, perturbative instability of de Sitter space and the related question of growth of perturbations during inflation has been widely discussed from many other and very different perspectives in the literature \cite{Myhrvold:1983hu,Ford:1984hs,Antoniadis:1986sb,Mukhanov:1996ak,Danielsson:2003wb,hep-th/0612138,astro-ph/0604488,0707.3377,hep-th/0605244,Garriga:2007zk,0801.1845,Urakawa:2009my,0912.2734,0912.1608,1005.3551,1008.1271,Urakawa:2010it,Byrnes:2010yc,1006.0035,1010.5327,1002.4214,1107.2712,1105.0418,Gerstenlauer:2011ti,Tanaka:2012wi,1302.3262,1302.6365,Tsamis:2013cka,1305.5705,1306.3846,1310.0367,1303.1068,Tanaka:2013caa,Frob:2014zka,1504.00894,1604.00390,1608.07237, Nacir:2016fzi, Finelli:2008zg, Finelli:2010sh}.

Another point is the similarity between the expressions (\ref{olbh}) and (\ref{olds}). In the present work we will be interested in a more precise calculation of this type of overlap.  We will focus on a horizon sized region, and consider the vacuum for this causal patch. As mentioned earlier, two soft gravitons will be emitted for every e-fold $\Delta t = 1/H$. Locally (within a single causal patch), the soft graviton mode can be gauged away and corresponds to a global symmetry. However, when comparing several causal patches (asymptotic observers), or between a transformed and un-transformed initial state, the symmetry is spontaneously broken, and the effect becomes physical. As noted in \cite{hep-th/0106109}, in pure de-Sitter the initial and final state should match up exactly, but when perturbations are considered that is no longer the case. Soft gravitons are then the Nambu-Goldstone bosons of the spontaneously broken symmetry. If we denote the charge generating the symmetry transformation by $Q$, then the state corresponding to adding a soft graviton to the vacuum can be obtained as $ e^{iQ}|0\rangle$. We will then show that if the state $|0'\rangle$ is obtained by acting on $|0\rangle$ with $e^{iQ}$ consecutively $\mathcal{N}$ times, corresponding to adding $\mathcal{N}$ soft gravitons, then we have for $\mathcal{N}\to \mathcal{N}_{dS}$
\beq
\langle 0| 0'\rangle\to 0~.
\eeq
Locally, adding a soft mode corresponds to a gauge transformation that does not fall off at infinity, and $Q$ is the conserved charge associated with the large gauge transformation. However, the symmetry is spontaneously broken between different local patches, and when comparing different observers, it becomes apparent that the transformation acts nontrivially on the state. This feature is also central to recent claims relating to the black hole information paradox \cite{1601.00921}. However, on small scales, where the symmetry remains unbroken, its effect is undetectable, which is at the core of the critique in \cite{1607.03120} (See also \cite{Averin:2016hhm} of another related discussion).

%On the other hand, a perturbative breakdown in inflationary spacetimes has been identified to occur on the analogue of the Page time (REF), which is given by the horizon size times the horizon entropy.  Incidentally, this is also the threshold for eternal inflation, hinting that indeed a more precise relation may exist between eternal inflation and the resolution to the information paradox.  In inflation, gravitons are constantly ripped out of the vacuum and stretched to superhorizon scales, which serves as the analogue of Hawking radiation in the black hole case. From a local point of view, these superhorizon modes look like soft modes, and can be gauged away by the de Sitter analogue of supertranslations.
%The main point of this paper is to construct the local charge related to the presence of soft superhorizon modes, and then see how it acts nontrivially on the vacuum state.  We will see that this changes the vacuum state in exactly the right way to reproduce both the Maldacena consistency relation and the one-loop consistency relations of GS. We will then use this to compute the timescale on which the average change in the local vacuum becomes large, and show that this is exactly the time scale of self-reproduction.

\subsection{Asymptotic Symmetries}

It has been known for over five decades that how you view the world depends on what type of person you are \cite{Bondi:1962px,Sachs:1962wk}.  By this, we are referring to the fact that the presence of a background metric breaks the full diffeomorphism symmetry of gravitational theories down to a much smaller subgroup, but which subgroup depends not only on the (asymptotic form of the) metric, but also on the asymptotic surface used to take the limiting behavior.  There has recently been a resurgence in this line of thinking in the case of Minkowski space \cite{1401.7026,1411.5745}, which, according to an observer at spacelike infinity possesses Poincar\'e invariance.  An observer at lightlike infinity, however, notices an enhanced symmetry group, referred to as the BvdBMS group, which also includes angle-dependent retardations known as supertranslations.  This has been argued to have physical consequences, as real observers more closely resemble those at lightlike infinity in many instances.  In this paper, we focus on the parallel of this discussion in de Sitter and inflationary spacetimes, which resemble de Sitter space up to slow roll corrections.  Though a spatial observer would conclude that de Sitter space possesses a $SO(4,1)$ analogue of Poincar\'e symmetry, a lightlike observer would notice the symmetry group to be greatly enhanced to that of diffeomorphisms on the future boundary.  These are the analogue of BvdBMS supertranslations noticed in \cite{1009.4730}, and, similar to the flat space case, possess physical consequences for observable quantities.  The relation between spatial diffeomorphisms and inflationary consistency relations was noted extensively in \cite{1203.6351,1304.5527} (see also f.ex.  \cite{1203.4595,Ghosh:2014kba,Kundu:2014gxa,McFadden:2014nta}), where the Noether charge was used to derive many of the previously known consistency relations, along with several novel extensions.  Our work extends their analysis in three ways:  first, we make the connection between the Noether charge and the Brown York charge explicit, which is important because the Brown York counterterms must be included in order to regulate both the ultraviolet and infrared divergences in the charge.  Second, we single out the explicit charge from the infinite class of possible charges that corresponds to the presence of a long mode, and show that using this one recovers not only tree level consistency conditions, but also loop corrections.  Lastly, we use the charge to encapsulate how the state evolves, and compute the overlap between states at two different times.  From this calculation we see explicitly that the overlap is driven to zero at the Page time of de Sitter space, signalling the breakdown of the effective description.

\subsection{Interpretation of the Brown-York charge}

The charge allows us to encapsulate the change in the state in terms of an operator acting on the vacuum.  It encodes the fact that the emission of a long wavelength graviton shifts the background coordinates by a spatial diffeomorphism.  In this paper we will mostly concern ourselves with the spin two part of the graviton $\g_{ij}$ rather than the scalar $\zeta$, for two reasons: during inflation, more scalar perturbations are emitted than tensors, making the timescale for tensors to become nonperturbative the lower bound on the full system.  Secondly, in de Sitter space the scalar mode is a pure gauge, so the tensor contribution will be the only one that is relevant in both cases.  This is illustrated in Fig \ref{jitter} below.  

\begin{centering}
\begin{figure*}[h]
\centering
\includegraphics[width=16cm]{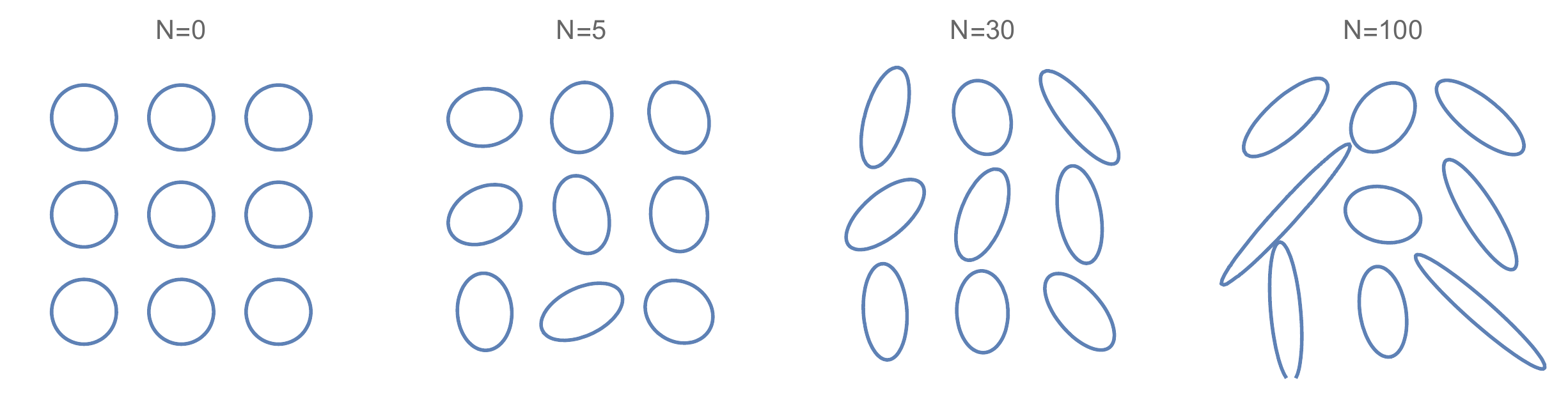}
\caption{A realization of the change in coordinates induced by the successive emission of long wavelength gravitons.  Circles here represent the horizon for an observer situated at their center.  Here $P_\gamma=.01$, so that each graviton emitted induces a $1\%$ deformation of the initial circles.  After the Page time, which in this case corresponds to 100 e-folds, the circles are so skewed that the initial global description of the system is no longer applicable.  To each local observer, however, this evolution is unobservable.}
\label{jitter}
\end{figure*}
\end{centering}

To a global observer, capable of keeping track of the entire inflated volume, the local horizons seem to jitter about stochastically, with an amplitude set by the power spectrum of gravitons.  Every local observer, however, will be blissfully unaware of the extreme squeezing they experience, as they are still in a locally flat spatial region.  Any ruler they could use to measure the amount of stretching in one direction relative to another would similarly adjust its length as they reorient it in various directions.  As the evolution of the system approaches the Page time, however, each circle is deformed by an $\mathcal{O}(1)$ amount, a simple consequence of the Brownian nature of the deformation process.  Once this happens the description in terms of the original coordinates becomes exceedingly poor. This is consistent with scale-dependent statistical anisotropies \cite{1104.0002}, the idea that de Sitter looks very anisotropic on large scales, but as one zooms in on smaller and smaller scales, it looks more and more isotropic.

\subsection{Gravitational Memory}

This effect is analogous to the phenomenon of gravitational memory \cite{zelpo}, where a passing gravitational wave induces a permanent displacement of neighboring freefalling observers.  Many connections between the memory effect and the asymptotic symmetry group have recently been elucidated, and our work extends this connection to the inflationary setting.  In flat spacetime, where the asymptotic symmetry group is the BvdBMS group, it was shown in \cite{1411.5745} that the passage of a gravitational wave out of a system induces a transformation that exactly corresponds to the gravitational memory effect that would be observed by a pair of idealized observers.  This provides a physical interpretation of the transformation by relating it to a measureable quantity that may soon be observable with LIGO \cite{1605.01415}, LISA \cite{1003.3486}, and pulsar timing arrays \cite{1410.3323}.  This was subsequently generalized to decelerating FLRW spacetimes in \cite{1602.02653} where it was found that because waves travelling in an expanding space have a contribution that lags behind the light cone, a readily measureable deviation from the flat space is induced.  This was extended to accelerating spacetimes and de Sitter space in \cite{1509.01296,1603.00151,1606.04894}, where it was shown that the memory effect is actually enhanced by a redshift factor over the flat space case.

\begin{centering}
\begin{figure*}[h]
\centering
\includegraphics[angle=-90,width=10cm]{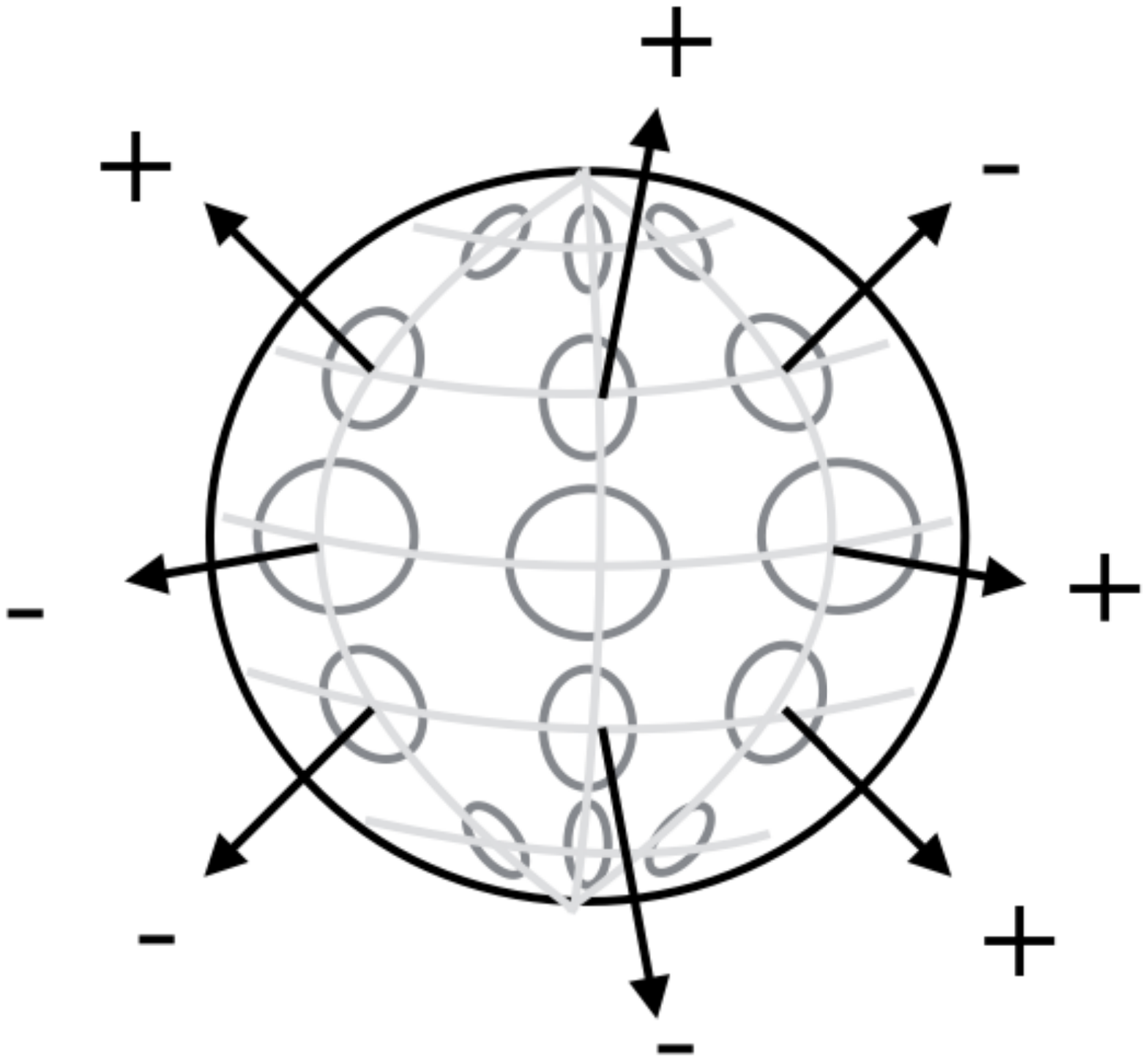}
\caption{An illustration of the memory effect.  As charges pass through the sphere, individual superconducting measurement devices register the induced vector potential as a pure gradient.  However, comparison across a network of detectors would allow to reconstruct the amount of charge that has left the system.}
\label{suss}
\end{figure*}
\end{centering}

A very insightful exposition of the relation between memory and the ASG is outlined in \cite{1507.02584}, which considers a network of superconductors arranged in a sphere surrounding a charge that subsequently escapes to infinity. The setup is arranged in such a way that the presence of the super conductors breaks the symmetry of the ground state.  Since superconductors directly measure the vector potential, this induces a phase in the detectors that to each looks like a pure gauge transformation.  However, when comparing the phases of the different detectors across the entire sphere it becomes clear that a physical event has taken place, as illustrated in Fig. \ref{suss}. This makes the connection with our setup most explicitly, as in the inflationary case the passage of a long wavelength graviton out of the system induces a change in the metric that to any one observer is pure gauge, but when comparing across many different observers becomes nontrivial as illustrated in Fig. \ref{pepsi}.  The difference in our scenario is that while it is relatively easy to do the global comparison for charges in flat space, in the inflationary case the presence of horizons prevents any comparison between two observers from ever taking place.

\subsection{Outline}

The paper is organized as follows:  In section \ref{2} we show how to construct the charge for a given long mode, and deal with various subtleties needed in its interpretation.  Section \ref{3} is devoted to using the charge to compute the overlap between states at two different times, which we show goes to zero after a Page time.  Section \ref{4} reproduces both the Maldacena consistency relation and the one loop correction to the two point function using our formalism, illustrating its utility.  We conclude in a discussion section, and leave various technical details to the appendices.

\section{Construction of the Charge}\label{2}
By analogy with the asymptotic symmetries of flat space, we argue that in an accelerating spacetime the quantum state does not have a unique vacuum, but instead has a similar large degeneracy of vacua that differ at subleading order.  These vacua correspond to the addition of an extremely long wavelength mode, and so a transition among the various states is naturally induced as the expansion causes soft modes to redshift to zero momentum.  Because these long wavelength modes do not carry energy, the states are degenerate, and the new state can rightfully be called a vacuum.  The first three subsections of this section are devoted to the construction of the charge associated with a long mode, and the last three subsections deal with its interpretation.

\subsection{Asymptotic Symmetries of Accelerating Spacetimes}

We consider spatially flat FLRW spacetime, described by the line element
\beq \label{FRWmetric}
ds^2=  -dt^2 + a^2(t)dx^2.
\eeq
We are interested in constructing the most general charge associated with the diffeomorphisms that leave this metric invariant at the asymptotic future boundary.  In Minkowski space, the causal structure of the spacetime played a major role, as the symmetry group of null infinity is greatly enhanced with respect to the symmetry group of spatial infinity.  Thus, instead of the simple asymptotic transformations that correspond to the Poincar\'e group, as found by \cite{Arnowitt:1959ah}, the group of asymptotic symmetries includes angle-dependent deformations of the null coordinate known as supertranslations (BvdBMS symmetries) \cite{Bondi:1962px}.  Further, it was argued \cite{1312.2229, 1401.7026, arXiv:1411.5745} that null infinity captures the relevant aspects of a physical detector, so that this enlarged group of transformations that act as symmetries to an asymptotic null observer have physical relevance.  

Accelerating FLRW spacetimes have distinct causal properties from Minkowski space due to the presence of their particle horizon, and so the form of the relevant asymptotic symmetries are very different in detail.  Nevertheless, the procedure can be repeated by analogy for this spacetime, and the full group of transformations that leaves null infinity invariant can be deduced.  The key distinction is that now null infinity coincides with spatial infinity as the conformal surface at which both null and timelike lines terminate.  In \cite{1009.4730} the residual diffeomorphisms were found (in de Sitter) to be just the group of spatial diffeomorphisms.  These play the same role as the BvdBMS symmetries in flat spacetime, in the sense that a soft graviton, as it redshifts to infinity, will act on the system as one of these transformations.  This can then be encapsulated by finding the precise diffeomorphism that corresponds to each soft graviton, and then tracking the change in the quantum state by using the Noether charge associated with that transformation.

When talking about FLRW space times the relevant gravitational degrees of freedom (the scalar $\zeta$ and the graviton $\gamma_{ij}$) can be extracted from the metric \ref{FRWmetric} as
\begin{eqnarray} \label{eq:metric}
ds^2=  -dt^2 +  a^2(t)e^{2\zeta} \left[e^{\g}\right]_{ij} dx^i dx^j
\end{eqnarray}

Here $\gamma_{ij}$ is transverse and traceless, and we have left out the lapse and shift, as they decay as powers of the scale factor, and so will be unimportant on the asymptotic boundary \cite{1203.4595}.  To construct the charge associated to the asymptotic symmetry group, we first perturb the background metric. The original work was done in synchronous gauge, as the asymptotic falloff behavior is guaranteed to only include power law decays \cite{Starobinsky:1982mr,1009.4730}.  However, to make contact with the cosmology literature, we prefer to work in comoving gauge.  More on the gauge fixing will be detailed below, where the results are shown to be independent of this choice.  The asymptotic symmetries act on this metric as \cite{1203.6351}
\beq \label{metricvariation}
\delta \left(e^{2\zeta} \left[e^{\gamma}\right]_{ij} \right)=\cL_\xi \left(e^{2\zeta} \left[e^{\gamma}\right]_{ij}\right)\eeq
where $\xi_i$ is an arbitrary spatial vector, and $\cL_\xi$ is the Lie derivative with respect to this vector.  

We will often be interested in the variation of the $\zeta$ and $\gamma$ fields under this transformation.  To do this we expand this defining equation to \cite{1203.6351} 
	\beq \label{delta zeta}
	2\delta\zeta\, h_{ij}+\delta h_{ij}=2\xi\cdot\partial\zeta h_{ij}+\xi\cdot\partial h_{ij}+\partial_i \xi^k h_{kj}+\partial_j \xi^k h_{ki}
	\eeq
where we defined $h_{ij}\equiv \left[ e^\g \right]_{ij}$ for ease of notation. We are able to extract the variation in $\zeta$ explicitly by multiplying the previous equation by the matrix $h^{-1}$, and taking the trace to negate all the dependence on the graviton.  This yields \cite{1203.6351} 
	\beq \label{zeta change}
	\delta\zeta=\xi\cdot\partial\zeta+\frac13\partial\cdot\xi,
	\eeq
which can be substituted back into \ref{metricvariation} to get an expression for the variation of the graviton that does not depend on $\zeta$.  However, this equation can only be solved perturbatively in $\gamma$, as there is no workable formula for the variation of the exponential of a matrix.  Fortunately, we will only be interested in the (next to) lowest order results, which are \cite{1203.6351} 
\beq \label{deltagij}
\delta\gamma_{ij}=\partial_i\xi_j+\partial_j\xi_i-\frac23\partial\cdot\xi\delta_{ij}+\xi\cdot\partial\gamma_{ij}+\mathcal{O}(\gamma^2).
\eeq

\subsection{Noether, Brown-York and Canonical Variables}\label{nbycv}

Because \ref{metricvariation} is an asymptotic symmetry of the theory, the Noether charge can be constructed in the usual way \cite{1304.5527}.  Though it is advantageous to use the Noether charge when it appears in commutators with fields, its main technical disadvantage is its volume divergence, discussed in \ref{V div}.  In the language of Brown and York \cite{Brown:1992br} the charge is a boundary term and local counterterms can be added to yield a finite quantity \cite{hep-th/9806087, hep-th/9811005}. We will need these counterterms in the next section, so we can regulate these volume divergences in the same line of what was done in \cite{hep-th/9902121, hep-th/0002230, 1009.4730}.  To see this, note that the Noether charge is given by

\begin{eqnarray} \label{QN}
Q_N(\xi)=\frac{1}{2} \int d^3 x\,  \left[ \Pi_\zeta \delta \zeta  + \Pi^{ij}_{\gamma} \delta \gamma_{ij} \right] +h.c.
\end{eqnarray}
where $\Pi_{\zeta,\gamma}=\delta \cL / \delta \left(\dot\zeta, {\dot\gamma_{ij}}\right)$ are the canonical momenta and $\delta \zeta, \delta \gamma_{ij}$ are the transformations of the canonical variables $\zeta$ and $\gamma_{ij}$ under the diffeomorphism $\xi$.  The Hermitian conjugate must be added to ensure that the charge is real when dealing with quantum operators.

Though this definition of the charge comes in handy when taking its commutator with the canonical fields, it has a major drawback in that both the conjugate momenta and variations of the fields are nearly impossible to write in closed form.  Because of this, we find it convenient to exploit the invariance of the form of the charge under field redefinitions to write
\beq 
Q_N=\frac12\int d^3x\, \tilde\Pi^{ij}\delta g_{ij}+h.c.
\eeq
where we simply use $g_{ij}=a^2 e^{2\zeta}e^{\gamma}_{ij}$ as our canonical variable.  Then, not only is the variation of this variable much easier to write down, but also the momentum can be written quite succinctly. We use the ADM form of the Einstein-Hilbert Lagrangian \cite{gr-qc/0405109}
\beq
\sqrt{g^{(4)}}R=\sqrt{g^{(4)}} \left(K_{ij}^2-K^2+R^{(3)} \right),
\eeq
where we have discarded total derivatives. Here, $K_{ij}$ is the extrinsic curvature, and the index-free version is traced with the full spatial metric. For simplicity we continue the derivation in pure de-Sitter where the momentum can be straightforwardly seen to be $2\tilde\Pi^{ij}= M_p^2 \sqrt{g}(K^{ij}-Kg^{ij})$, so that the charge is 
\beq \label{QN2}
Q_N=\frac{M_p^2}{2} \int d^3x\sqrt{g} (K^{ij}-Kg^{ij})\nabla_{(i}\xi_{j)}.
\eeq
We stress that this charge is defined exactly for the full theory. 
This form makes it very easy to see the equivalence between the Noether charge as defined in \ref{QN} and the Brown-York charge \cite{Brown:1992br}.  If the spatial integration is over a 3-dimensional hypersurface $\Sigma$, this can be integrated by parts to yield a boundary term (also using the momentum constraint, $\nabla_i \Pi^{ij}=0$)
 associated with the 2-dimensional boundary $\partial \Sigma$
\begin{eqnarray}
Q_{BY}= \int_{\partial \Sigma} d^2 x \sqrt{ \sigma} n^i \xi^j T_{ij}^{BY}
\end{eqnarray}
where $\sigma$ is the induced metric in $\partial \Sigma$, and $n^i$ is a vector tangent to $\Sigma$ and normal to $\partial \Sigma$.  This is the Brown-York charge with the associated Brown-York tensor

\begin{equation}
T^{BY}_{ij}= M_p^2 \left( K_{ij} - K h_{ij} \right).
\end{equation}

\subsection{Determination of the Diffeomorphism}
Now that we have an expression for the Noether charge for a generic diffeomorphism, we can single out the unique charge to use for a precise soft mode by matching the effect the charge has on short wavelength modes encoded in the commutator $[Q,\mathcal{O}]=-i\delta\mathcal{O}$. A similar procedure was already done in \cite{1304.5527}, but there generic charges were used to prove novel consistency relations, as opposed to our work, which aims to use a specific charge to find the effect on the state and track the time evolution of the system. We set the convention by considering scalar modes first, and then show that it acts on tensor modes in the same way.  We first argue that we can neglect the second term in \ref{zeta change} for short wavelength modes. We consider that $\xi $ only encodes the information associated with soft modes (which have become classical) such that even at higher orders no hard modes show up. Then, by taking the Fourier transform, since $\xi$ only has support for extremely small momentum, the effect on a large momentum mode will vanish.  Thus, the second term can be neglected, and we can use this to find an explicit expression for the curvature perturbation in the new vacuum.  From (\ref{zeta change}), (\ref{QN}) we have
	\beq
	[Q,\zeta(x)]=-i\xi\cdot\partial\zeta(x).
	\eeq
	Then, successive applications of the commutator become multiple actions of the operator $\xi\cdot\partial$ on the observable, so that
	\beq\label{expQ}
	e^{-iQ}\zeta(x)e^{iQ}=e^{-i[Q,\cdot]}\zeta(x)=e^{-\xi\cdot\partial}\zeta(x)=\zeta(x-\xi)~.
	\eeq
It is important to interpret the exponential in the third equality as the differential operators acting only on the field, and not on themselves.  This expression implies that the action of the unitary transformation induced by the Noether charge is just to shift the coordinates by the diffeomorphism explicit in the charge.  In some way, this expression is patent, as this is exactly what the charge was constructed to do.  On the other hand, this proves that a shift of the coordinates can be thought of as a change of the state of the system, and it provides the exact transformation that can be used to compute corrections to correlation functions order by order, which is what allows us to recover the known consistency relations and loop corrections in section \ref{4}.  This allows us to write the diffeomorphism explicitly as
\beq
\xi_i=\left(\delta_{ij}- e^{\gamma^L/2}{}_{ij}\right)x^j,
\eeq
where $\gamma^L_{ij}$ denotes the soft graviton.	
Note that this can be extended to an arbitrary number of curvature perturbations, evaluated at either coincident or distinct spacetime points.  Extending this result to gravitons is not as trivial, since we cannot write the explicit expression for how the graviton changes, but we can exploit the fact that $e^{-iQ} f(g_{ij}(x))e^{iQ}=f(g_{ij}(x-\xi))$ for any function $f$.  Taking $f$ to be a logarithm, we get that $e^{-iQ}(2\zeta(x)\delta_{ij}+\gamma_{ij}(x))e^{iQ}=2\zeta(x-\xi)\delta_{ij}+\gamma_{ij}(x-\xi)$.  Using the transformation properties of $\zeta$ we have already uncovered, we have proved that gravitons transform in the same way.  This can now be extended to an arbitrary function of these two variables.

Note that equivalently we can think of $Q$ as the operator corresponding to a small linearised symmetry transformation of the form $x\to x-\xi^L$, with $\xi^L$ just being given by the linear term
\beq
\xi^L_i =-\frac{1}{2}\gamma^L_{ij}x^j  ~,
\eeq	
but now letting the differential operators act on them self (but assuming still that the long mode, $\gamma^L$, is constant). In that case 
\beq\label{expQ2}
	e^{-iQ}\zeta(x)e^{iQ}=e^{-i[Q,\cdot]}\zeta(x)=e^{-\xi^L\cdot\partial}\zeta(x)=\zeta(x e^{\gamma^L/2})~.
	\eeq
The equivalence of the two approaches follows from the fact that a large transformation can be generated by acting repeatedly with a small symmetry transformation, and demonstrates why the non-linear extension of the linear Gaussian perturbations on super-horizon scales take the exponential form.

\subsection{Goldstone Charges and Volume Divergence} \label{V div}

Since we are advocating thinking of spatial diffeomorphisms as a change in the vacuum state, we must necessarily address the fact that the charge we constructed is formally infinite.  This is a generic feature of charges constructed from spontaneously broken symmetries \cite{fabpis}. Since the initial vacuum selects from an infinite class of metrics with the same de Sitter isometries, the soft gravitons act as Goldstone modes, and the charge is proportional to the spatial volume. 
This may be best understood by taking a simple analogy: that of a ferromagnet. In this example, the ground state of the system spontaneously breaks rotational invariance by selecting out a preferred direction along which all the microscopic magnets align.  These can be rotated to a new direction, but if the ferromagnet is infinite it would take an infinite amount of work, and the charge would diverge. This limit corresponds to adding a single long wavelength ``roton'' and, because the charge is infinite, the two states have exactly zero overlap.  This indicates that the Hilbert spaces of the two ground states are exactly separable.  This is no more mysterious than the fact that in a Fock space, one particle states are orthogonal to zero particle states: since the new state is related to the old by the addition of a single (long wavelength) particle, they are disconnected.  Manipulation of the state directly leads to divergences \cite{1001.5212, 1210.7792}, but the calculation of any correlator can be done explicitly, because when the exponential is expanded the charge only appears in commutators \cite{1210.7792 ,1304.5527}.  Because the canonical commutation relations involve delta functions, any reference to the volume integration is nullified and all commutators become well defined.  

However, if we restrict to any finite region, the manipulations of the state do not have to be treated as purely formal, as all expressions will be finite.  The price we pay for this is losing the property that the charge is conserved: since we deal with a finite region, charge can leak in or out of the region under consideration corresponding to the emission of a soft mode.  In practice, we renormalise the volume divergences by adding adequate counterterms on the boundary, and picking the term which is independent of the finite volume we choose. This is done in appendix \ref{Renormalization} by considering 
\begin{eqnarray} \label{ActionCounterTerms}
S &= & M_p^2 \left[ \int d^4 x \sqrt{-g^{(4)}} \frac{R}{2} + \int_{\cI} d^3 x \sqrt{-g^{(3)}} \left( -K[g]+ \right. \right. \nn \\ && \underbrace{ \left. \left. - 2 c_1 H + \frac{c_2}{2} R[g] +  c_3 R[g]^2+ c_4 R[g]_{ij} R[g]^{ij}   \right) \right]}_\text{Counter-terms} \, .
\end{eqnarray}
Interestingly, the UV divergences appearing in the overlap computation in section \ref{Overlap} are also renormalized by this procedure.

\subsection{Global vs Local}

There are two different states we will be interested in, a distinction between which needs to be made.  

The first is the transformation of the vacuum state under the symmetry transformation associated with a particular long wavelength mode encoded in the diffeomorphism $\xi$: \begin{eqnarray}
|0'\rangle=e^{iQ(\gamma_L)}|0\rangle \,.
\end{eqnarray} 
This state automatically shifts the coordinates of all quantities in the correlators used, and so represents a shift in the background for all subhorizon correlators.  In this way, the state itself changes in such a manner that is imperceptible to a local observer.  Said another way, any long wavelength configuration can be viewed as a shift in the background coordinates that coincides with the asymptotic isometries of the spacetime, and, since there is no preferred metric in the class of asymptotic FLRW's, any local measurement is unable to distinguish such a shift.  In this way, our method of calculating closely resembles the Schr\"odinger picture, where the time evolution is encapsulated in the state, rather than the Heisenberg or interaction picture.  This allows us to directly compare states at two different times, in order to characterise the time evolution in terms of the mutual overlap between two different states, as was done in \cite{hep-th/0703116} in the case of black holes.

\begin{centering}
\begin{figure*}[h]
\centering
\includegraphics[angle=-90,width=8cm]{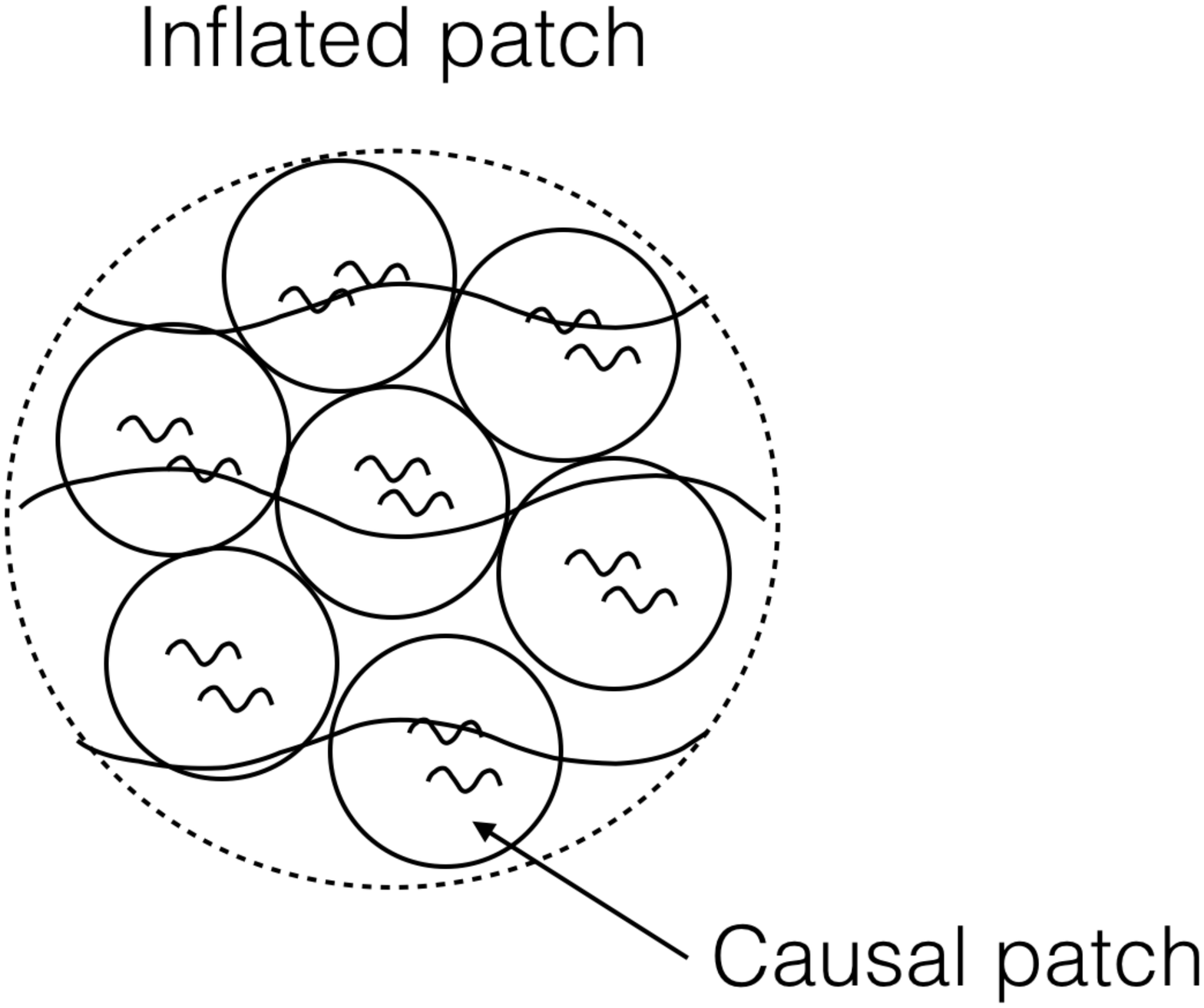}
\caption{\label{Inflated patch}A pictorial representation of the influence of long wavelength modes.  Each observer in the small circles is oblivious to the soft modes, and observes all correlators in a locally flat frame.  The long wavelength modes induce a variance in the observed quantities over many different patches, however, and the double expectation value (defined as in \cite{1005.1056}) captures this phenomena.}
\label{pepsi}
\end{figure*}
\end{centering}

The second state of interest is the expectation value of the first state. Here we choose an arbitrarily large volume on which we take the expectation value of the long modes themselves, and compute the expected deviation of given observables in this region. We denote this as $\langle\langle 0'| \mathcal{O}|0'\rangle\rangle$.  This is emphatically not what would result from any measurement that could be performed by any local observer, but instead captures the variance of the measurements of the many different observers in the large volume chosen. This quantity measures the global departure from the original state, as if this becomes large the state will in no way resemble the initially prepared FLRW background, even though any local observer will be oblivious to this fact. In appendix \ref{Ergodicity} we show that this average can also be seen as a continuous transformation of the state by the symmetry transformation.

\subsection{Gauge Invariance}
A brief note on the gauge fixing: cosmologists typically use the comoving gauge, where the lapse and the shift are constrained to be functions of the inflaton fluctuations, whereas in defining the Brown-York tensor the synchronous gauge is used, so as to simplify the form of the asymptotic symmetry group.  We must address this mismatch as we are to compare results between the two gauges.  Fortunately, as is well known, in gravitational theories the Noether charge is a gauge invariant quantity, with a few provisos.  Firstly, as was found in \cite{Crnkovic:1986ex, Lee:1990nz}, this is because the symplectic form that can be derived from the Noether charge has degenerate directions, which are associated with the gauge symmetries of the theory.  When restricted to the constraint surface defined by the space of fields satisfying the equations of motion, the charge is independent of the gauge.  Ordinarily, this also implies that the charge associated with a diffeomorphism is purely a boundary term, which vanishes identically if suitable falloff conditions can be imposed.  However, for large gauge transformations the charge remains nontrivial \cite{hep-th/0205072,1604.07764}, which is the reason why our procedure yields interesting results.  However, in going between the comoving and synchronous gauges, a small diffeomorphism can be used if we restrict the fluctuations used to define the coordinate transformations to a finite region.  Thus, the charges will be equal in the two different gauges, and we are free to use the most convenient gauge for our purposes.
	
\section{Overlap} \label{Overlap}\label{3}

In the previous section we detailed the construction of the charge associated with a given finite volume at the future infinity ($I^+$) of de-Sitter.  Now we use the charge to track the evolution of the state itself through the continued emission of long wavelength modes.  To do this, we suppose the existence of a meta-observer capable of comparing observables at different Hubble volumes (at $I^+$ or in a subsequent stage after de-Sitter).  These are the ``calculables'' from \cite{hep-th/0106109}, which were argued to be the most appropriate encapsulation of physics in de Sitter space. In analogy with the black hole setting \cite{hep-th/0703116}, we show that the states at two different times become almost orthogonal to each other when the amount of emitted gravitons becomes comparable to the number of horizon degrees of freedom, which occurs when the Page time \cite{hep-th/9305040} of de Sitter has elapsed.
In this computation we only consider the charge associated to soft gravitons and neglect slow-roll corrections in the mode functions. Although our results are valid for pure de Sitter, the generalization to single-field inflation should be straightforward by both including the charge associated with scalars and mixed components, and by considering slow-roll corrections. In this case, we expect the same parametric result, modified by slow-roll factors.

In computing the overlap, several subtleties show up, which we summarize here:

\begin{itemize} 
	\item{{\it Regularization:} 
		
		Since the charge $Q$ corresponds to a spontaneously broken symmetry (the presence of a long graviton selects a particular spatial slicing) it is actually an ill-defined quantity (see section \ref{V div}).  This is a regular occurrence in theories with spontaneously broken symmetries \cite{1001.5212}.  It is solely due to the fact that the charge is integrated over an infinite region of space, and so becomes finite whenever we restrict to a finite region.  This IR regularization will cause the charge to lose its conservation property, as it would then be possible for charge to leak into or out of the bounded region under consideration.  However, if we average over many regions, this is equivalent to taking the volume to be large even compared to the soft mode.  In the big region, the mode will not act as a shift of the background coordinates, but the state will still be the original vacuum.  However, the variance of the charge will grow, due to its nonconservation in each small region.
		
There may seem to be a slight ambiguity here, as the charge scales extrinsically with the volume chosen, and so would be sensitive to whether we choose to consider the charge within one Hubble patch, or some other volume that is off by an $\mathcal{O}(1)$ factor.  In fact, we will show in appendix \ref{Renormalization} that the renormalization procedure removes all dependence on this arbitrary choice.  While divergent terms are sensitive to this number, the renormalized finite terms are not.
} 
	\item{{\it UV renormalization}: 
		
		In addition to the infrared divergence, the expectation value of the charge has the usual ultraviolet divergences, which we treat using standard renormalization techniques in appendix \ref{Renormalization}.  Extracting the finite piece introduces an ambiguity as usual, but this does not affect the qualitative assessment that the overlap between states vanishes after a Page time.  Indeed, since the Page time is the amount of time it takes for the degrees of freedom emitted to rival those on the de Sitter horizon, characterising this quantity very precisely would require some ad-hoc definition. }
	\item{{\it Perturbativity:}
		
		In lieu of being able to explicitly compute the overlap of the states at two different times to all orders, we are in search for a useful diagnostic to determine the timescale over which the states become orthogonal.  For this we use perturbative methods, both by expanding the exponential containing the charge, and the charge itself in terms of the fields.  For the former, we must be careful to pay attention to partial resummations that can be performed, as they will have a significant impact on our conclusions.  In particular, terms that are linear in the charge resum to a pure phase, leaving the quadratic terms as the dominant contribution.  We show that the expectation value of the square of the charge becomes $\mathcal{O}(1)$ after a Page time has elapsed, in line with our expectations.  In appendix \ref{Resummation} we argue that including higher order corrections will cause these terms to assemble into a function that decreases to 0 on this timescale.}
	\item{{\it Contact terms:}
		
		Evaluation of the overlap involves numerous contact terms, in which propagators are evaluated at vanishing separation. We show in the appendix \ref{Contact terms} that we do not have to worry about these, as they always assemble into an irrelevant pure phase. Thus, we focus our attention on the remaining parts of the quadratic expectation value, which are not evaluated at coincident points.}
	\item{{\it Quadratic terms:} 
		
		Even among non-contact terms not all relevant. The terms in $Q$ quadratic in $\gamma$ are forced to have the same momenta on-shell. Thus, due to our separation of scales between the hard and the soft modes there will be no contribution from quadratic terms.}
\end{itemize}

After all these subtleties, the leading effect becomes $\langle Q_3^2\rangle_\text{NC} $, where $Q_3$ is cubic in the fields and NC denotes the non-contact terms in the charge.  When this quantity becomes of order one the overlap between states goes to zero and the vacuum description in the big volume breaks down.  This corresponds exactly to the expectation that the effective field theory description breaks down after one Page time.

\subsection{Change in the local vacuum}
	
As we saw in section \ref{2} the charge acts on correlators as a large gauge transformation which is physically equivalent to the creation of soft modes (gravitons or scalars). This transformation captures the local evolution of the vacuum where modes become superhorizon as time evolves.

Consider the initial vacuum state denoted by $\left|0\right>$. In the presence of a long wavelength field the state undergoes the symmetry transformation
\begin{eqnarray} \label{shifted state}
 \left|0'\right> = e^{i Q} \left|0\right>.
\end{eqnarray}
In order to compare how much the two states differ we compute their overlap (see Fig. \ref{pepsi}). 
\begin{eqnarray}
\left<0'|0\right> = \left<0 \right| e^{iQ} \left|0 \right> = \left<0\right|\left(1+iQ-\frac{Q^2}{2} + \cO \left(Q^3\right) \right) \left|0\right>.
\end{eqnarray}
When this matrix element becomes much smaller than one it means that our standard description of the vacuum is no longer accurate. 
As we explained in the beginning of this section, the contact terms sum up to a physically irrelevant phase (see appendix \ref{Contact terms}) while the quadratic part of the charge vanishes in the finite region we are integrating. Therefore, the overlap becomes
\begin{eqnarray}
\left|\left<0|0'\right>\right|= \left|1-\frac{1}{2}\left<Q_3^2\right>_\text{NC} +\cdots \right|
\end{eqnarray}
where $\cdots$ are terms suppressed by powers of $(H/M_p)^2$ and the subscript NC denotes non-contact terms. 

Following section \ref{nbycv} we consider the charge associated with a volume $V$ in a space-like surface at the future infinity of de-Sitter space. We focus on the pure graviton part of the charge, because it is relevant for both de Sitter and slow roll:
\begin{eqnarray}
Q= \frac{1}{2}  \int_V d^3x \,  \Pi^{ij} \delta \g_{ij}  + h.c.
\end{eqnarray}
where $\Pi^{ij}$ is the graviton canonical momenta and $\delta \g_{ij}$ is the change in the hard graviton due to the diffeomorphism (c.f. \ref{deltagij}), which to lowest order is given by
\beq
\delta\gamma_{ij}=-\gamma^L_{ij}-\frac12\gamma^L_{ab}x^a\partial^b\gamma_{ij} + \cO(\g^L{}^2,\partial\g^L, \g^2).
\eeq
Here $\g^L_{ij}= \g^L_{ij}(t) e^{-i k_L\cdot  \x}$ is the long wavelength graviton, whose wavelength is assumed to be much larger than the size of the volume $V$. For simplicity we assume $V$ to be a sphere of radius $b/(aH)$ where $b$ is an arbitrary coefficient parameterizing the size of the sphere. 

For standard Einstein gravity, because the cubic interactions only contain spatial derivatives, the canonical momentum does not have quadratic corrections \cite{astro-ph/0210603} and is given by
\begin{eqnarray} \label{Canonical Momentum}
\Pi_{ij}= \frac{\delta (\sqrt{-g} \cL)}{ \delta \dot{\g}_{ij}}= \frac{   a^3 M_p^2 }{4} \dot{\g}_{ij} + \cO (\g^3).
\end{eqnarray}
Therefore, using eqs. \ref{deltagij} and \ref{Canonical Momentum}, the cubic part of the charge becomes
\begin{eqnarray} \label{CubicCharge}
Q_3(t)&=& -\frac{a^3 M_p^2}{16}  \int_V d^3 x \, \dot{\g}_{ij}  \g^L_{ab} x^b \p^a \g_{ij}+h.c. \\
 &=& -\label{CubicCharge_kspace} \frac{a^3 M_p^2}{8}  \text{Re}\int_V d^3 k \, \dot{\g}_{ij}  (-k) 2 D_L \g_{ij} (k).
\end{eqnarray}
where $D_L=\frac12\g_{ab}^L \partial_{k_a} k^b $.  In the second line we have treated the long mode $\g^L$ as constant inside $V$, neglecting corrections which will be proportional to $k_L x$.  In section \ref{4} we will show that this charge allows us to reproduce both the graviton three point consistency relation and the loop corrections to the scalar two point function, but now we use it to compute the overlap between the two different states.

From before, the leading contribution to the overlap is
\begin{eqnarray}
\left<Q_3^2(t)\right>_\text{NC}= \left(\frac{a^3 M_p^2 }{8}  \right)^2 \int d^3 k d^3 q  \Big< \text{Re}[\dot{\g}_{ij} (-k)  2D_L(k) \g_{ij} (k)]\, \text{Re}[\dot{\g}_{mn} (-q)  2D_L(q) \g_{mn} (q)]  \Big>_\text{NC}.
\end{eqnarray}

We will spare the reader from the many intermediate steps in evaluating this expression, and instead list the most important details of the procedure.  Wick contraction will give two terms, one with time and momentum derivatives segregated, and one with them mixed\footnote{Note that we do not need to consider gauge fixing ghost contributions, since we are working in a fully gauge fixed formalism in which covariance on the other hand is not manifest. This is similar to working in ``physical gauge" versus ``covariant gauge".} .  Both these will be proportional to two powers of the same delta function.
The first will reduce the expression to a single momentum integral, and the second will yield the integration volume factor $V$.  Moreover, we can also make the replacement $2D_L f = \g^L_{ab}\hat k^a\hat k^b f'$, $f'=k\partial_k f$ for all quantities we take this derivative of, as a consequence of the transverseness and tracelessness of gravitons.   Therefore, all terms will have the same tensor structure and angular dependence, which can be integrated over using $\int d\Omega\langle \gamma^L_{ab}\g^L_{cd}\rangle \hat k_a\hat k_b\hat k_c\hat k_d=32\pi/15\langle \g^L \g^L \rangle$.  After all this, the correlator becomes
\begin{eqnarray} \label{CubicCharge_preint}
\left<Q_3^2(t)\right>_\text{NC}= \frac{8\pi}{15} V a^6 M_p^4 \langle\g^L\g^L\rangle \int_{aH/b}^{a \Lambda} dk k^2 |\dg|^2 \left| \g'_k\right|^2
\end{eqnarray}
where the final momentum integral goes from $aH/b$ to the UV cutoff $a \Lambda$.  This integral is both UV and IR divergent, and so a finite answer must be extracted from a renormalization procedure.  This is done in detail in appendix \ref{Renormalization}, but here we quote the result: the integral becomes a pure number, yielding
\begin{eqnarray}
\left<Q_3^2(t)\right>_\text{NC}= \frac{8 \pi^2}{45} \left<\g^L \g^L\right>. 
\end{eqnarray}
The previous equation tells us that the change in the vacuum is proportional to the 2-point function of the long mode times a coefficient of order 1. Recall, however, that $\g^L$ is a particular soft mode of size larger than the volume $V$. We now would like to average the overlap over many spheres of size V inside a bigger region, which we could think of as the total inflated patch, where the variations in the long mode become relevant (see fig. \ref{Inflated patch}). We denote this operation by a double averaging, namely, $\left< \left< \cdots \right> \right> $ similar to the procedure of \cite{1005.1056}. In appendix \ref{Ergodicity} we show that this procedure is equivalent to applying the charge continuously over a given range of soft modes, hence, it suggests the ergodicity of the process. 
The averaging of $\left< \g^2 (x)\right>$ is the variance of the long mode in the big volume and is given by \cite{1005.1056}
\begin{eqnarray}
\left< \g^2 (x,t)\right>= - \frac{H^2}{2\pi^2M_p^2}\log\left(\frac{\Lambda_\text{IR}}{a(t)H}\right)
\end{eqnarray}
where $\Lambda_\text{IR}=H$ is the largest comoving scale, the initial scale factor having been set to 1. Therefore, 
\begin{eqnarray} \label{Double Average of Q}
\left< \left< Q_3^2 \right> \right> \simeq   \frac{4}{45} \frac{H^2}{M_p^2}\log\left(a(t)\right)
\end{eqnarray}
which means, that the average overlap between the two vacuum goes to zero on the time scale
\begin{eqnarray}
\frac{H^2}{M_p^2} \log\left(a(t)\right) \simeq 1,
\end{eqnarray}
which, using $\log(a(t))= Ht$, is exactly the equivalent of the Page time for the de Sitter: $t= M_p^2/H^3$.

\section{Recovering consistency relations}\label{4}

In this section we use the charge constructed in \ref{2} to recover several known consistency conditions for inflationary correlators.  We begin with the consistency relation for gravitons, and afterwards show that the one loop correction to the scalar correlator is reproduced as well.  This not only convincingly shows that the viewpoint we adopt in this paper is accurate, but also highlights the utility of this line of reasoning, as the effort required to reproduce these results is rather minimal.

\subsection{Graviton 3-point function}

We start by the derivation of the graviton 3-point function in the squeezed limit
\begin{eqnarray}
\left< \g^{s_1}_{q_1} \g^{s_2}_{q_2} \g^{s_3}_{q_3} \right>
\end{eqnarray}
where $s_i$ are the polarizations and $q_1 \ll q_2, q_3$. 

In the standard picture, this correlator is zero at tree-level and one needs to go to first order in the in-in formalism and insert one interaction Hamiltonian. 
In the charge picture the vacuum itself is changed by the symmetry transformation in such a way that it already contains a long mode entangled with the short scale physics.
Then, in the new vacuum
\begin{eqnarray} \label{g3pf}
 \left< 0'\left| \g^{s_1}_{q_1} \g^{s_2}_{q_2} \g^{s_3}_{q_3} \right| 0'\right> = \left< 0\left|e^{-iQ} \g^{s_1}_{q_1} \g^{s_2}_{q_2} \g^{s_3}_{q_3} e^{iQ}\right| 0\right> \equiv \left< \g^{s_1}_{q_1} \g^{s_2}_{q_2} \g^{s_3}_{q_3} \right>_Q  
\end{eqnarray}
We proceed by expanding \ref{g3pf} in powers of $Q$
\begin{eqnarray}
\left< 0\left|e^{-iQ} \g^{s_1}_{q_1} \g^{s_2}_{q_2} \g^{s_3}_{q_3} e^{iQ}\right| 0\right> = \left< \g^{s_1}_{q_1} \g^{s_2}_{q_2} \g^{s_3}_{q_3} \right> - i \left< \left[Q, \g^{s_1}_{q_1} \g^{s_2}_{q_2} \g^{s_3}_{q_3} \right] \right> + \cO(Q^2).
\end{eqnarray}
As we referred before, the first term on the right hand side is zero and the second piece is the leading contribution. Using eq. (\ref{CubicCharge}) we have
\begin{eqnarray}
\left< \left[Q, \g^{s_1}_{q_1} \g^{s_2}_{q_2} \g^{s_3}_{q_3} \right] \right> =-\frac{a^3 M_p^2}{8} \left<  \left[  \int_V d^3 x \, \dot{\g_{ij}}(x)  \g^L_{ab} x^b \p^a \g_{ij}(x), \g^{s_1}_{q_1} \g^{s_2}_{q_2} \g^{s_3}_{q_3} \right] \right>\,.
\end{eqnarray}
Given our volume restriction the soft modes contract with each other and can be pulled out of the commutator, leading to\footnote{Note that only the diffeomorphism associated with a long mode of momentum $q_1$ will give a non-zero result. The correct way to think about it is by considering an integration of the charge over diffeomorphisms which would be equivalent to integrating over soft modes.}
\begin{eqnarray}
\left< \left[Q, \g^{s_1}_{q_1} \g^{s_2}_{q_2} \g^{s_3}_{q_3} \right] \right>  \xrightarrow[q_1\rightarrow0]{} -\frac{a^3 M_p^2}{8} \left< \g^L_{ab} \g^{s_1}_{q_1} \right>  \left<  \left[  \int_V d^3 x \, \dot{\g_{ij}}(x)  x^b \p^a \g_{ij}(x), \g^{s_2}_{q_2} \g^{s_3}_{q_3} \right] \right>.
\end{eqnarray}
After Fourier transforming, contact terms do not contribute and contractions between short modes give a factor of 2 leading to
\begin{eqnarray}
\left< \left[Q, \g^{s_1}_{q_1} \g^{s_2}_{q_2} \g^{s_3}_{q_3} \right] \right> = -\frac{a^3 M_p^2}{4}\, \e^{s_1}_{ab}(q_1)  \hat{q}^a_2 \hat{q}^b_2 \,\left< \g^L \g^L\right>' \,\e^{s_2}_{ij} \e^{s_3}_{ij} \,(2i)\text{Im} \left[  \dot{\g}_{q_2}  \g'_{q_3}  \g_{q_2}^* \g_{q_3}^* \right] \,\delta^{(3)}(q_2+q_3)
\end{eqnarray}
Inserting the mode functions (eq. \ref{modfunc}) and using that $\e^{s_2}_{ij} \e^{s_3}_{ij} = 2 \delta_{s_2,s_3}$  we get, in the de Sitter limit, and after the short modes become superhorizon, 
\begin{eqnarray}
\left< \left[Q, \g^{s_1}_{q_1} \g^{s_2}_{q_2} \g^{s_3}_{q_3} \right] \right> &=&  \frac{3}{2} i  \e^{s_1}_{ab}(q_1) \,\hat{q}^a_2 \hat{q}^b_2 \, \left< \g^L \g^L\right>' \left<\g^{s_2}_{q_2}\g^{s_3}_{q_3} \right>' \delta_{s_2,s_3} \delta^{(3)}(q_2+q_3).
\end{eqnarray}
Therefore,
\begin{eqnarray}
\left<  \g^{s_1}_{q_1} \g^{s_2}_{q_2} \g^{s_3}_{q_3} \right>_Q &=& \frac{3}{2} \, \e^{s_1}_{ab}(q_1) \hat{q}^a_2 \hat{q}^b_2 \, \left< \g^L \g^L\right>' \left<\g^{s_2}_{q_2}\g^{s_3}_{q_3} \right>' \delta_{s_2,s_3} \delta^{(3)}(q_2+q_3).
\end{eqnarray}
which agrees with the result found in \cite{astro-ph/0210603}.

\subsection{One loop Correction}

We also demonstrate that the change in state is able to reproduce loop corrections to the correlators calculated in the literature.  We choose to illustrate this for the two point scalar correlator, as computed in \cite{1005.1056}.  For this we need the scalar part of the charge
\beq
Q=\int d^3x \, \Pi_\zeta\left(D_L\zeta+\frac13\nabla\cdot\xi\right).
\eeq
The novelty here is that now we need to keep higher orders in the long graviton mode, so that the differential operator is $D_L=-(\gamma_L/2+\gamma_L^2/8)_{ij}x^i\partial_j+\dots\rightarrow(\gamma_L/2+\gamma_L^2/8)_{ij}  \partial_{k_i}  k_j$.  As before, if we restrict our attention to the influence on short wavelength modes, the second term in the charge can be neglected, and we have $[Q,\zeta_k]=-iD_L\zeta_k$.  Then, the mode function can be expanded as
\beq
e^{-iQ}\zeta_k e^{iQ}=\zeta_k-D_L\zeta_k+\frac12 D_L^2\zeta_k+\dots
\eeq
The differential operator has some $k$ dependence, and so acts nontrivially\footnote{It is important to remember that by the definition in eq. (\ref{expQ}), $D_L$ cannot act on itself. However, we could equivalently have used the linearised form of $Q$ in eq. (\ref{expQ2}) in terms of $\xi^L$ and then letting $D_L$ act on itself. We checked that the two approaches yield the same result.}.  To quadratic order in $\g_L$,  
\beq
D_L^2 f_k=\frac14\bigg(\delta_1{}^2 \Big((k \cdot \partial_k)^2-2k \cdot \partial_k\Big) + 3 \delta_2 (k\cdot \partial_k) + \left< (\g_{Lij})^2\right>\bigg)f_k
\eeq
Where we have defined
\beq
\delta_1{}^2=\frac{1}{4\pi}\int d\Omega\langle\gamma_{Lij}\gamma_{Lkl}\rangle\hat k_i\hat k_j\hat k_k\hat k_l=\frac{8}{15}\langle\gamma_L^2\rangle,\quad\delta_2=\frac{1}{4\pi}\int d\Omega\langle\gamma_L^2{}_{ij}\rangle\hat k_i\hat k_j=\frac{4}{3}\langle\gamma_L^2\rangle
\eeq
Then the full expression for the mode function becomes
\beq
e^{-iQ}\zeta_k e^{iQ}=\left(1+\frac12\delta_1k\cdot\partial_k+\frac14(\delta_2-\delta_1{}^2)k\cdot\partial_k+\frac18\delta_1{}^2(k\cdot\partial_k)^2\right)\zeta_k
\eeq
and
\beq
\langle\zeta_k\zeta_q\rangle_Q=\left(1+\frac14\delta_1{}^2k\cdot\partial_kq\cdot\partial_q+2\left(\frac14(\delta_2-\delta_1{}^2)k\cdot\partial_k+\frac18\delta_1{}^2(k\cdot\partial_k)^2\right)\right)\langle\zeta_k\zeta_q\rangle+\dots
\eeq
This yields
\beq
\langle\zeta\zeta\rangle_Q=\left(1+\langle\gamma_L^2\rangle\left(\frac25k \cdot \partial_k+\frac{4}{15}(k \cdot \partial_k)^2\right)\right)\langle\zeta\zeta\rangle+\dots
\eeq
In agreement with \cite{1005.1056}.  If the running of the two point function is negligible this can be further reduced to the simple expression
\beq
\langle\zeta\zeta\rangle_Q=\left(1+\frac{(4-n_s)(1-n_s)}{15}\langle\gamma_L^2\rangle\right)\langle\zeta\zeta\rangle+\dots
\eeq
which is in fact slow roll suppressed, and vanishes in pure de Sitter space.
\section{Discussion}

In this paper we have demonstrated that the presence of a long mode, which is normally treated as a shift in the coordinates of all quantities to be evaluated, can equally well be thought of as changing the quantum state of the system.  We found the unique shift associated with a particular long wavelength mode, and explicitly showed that calculations in this new state reproduce standard results quite readily.  Furthermore, having this method of calculation in our arsenal allowed us to address aspects of the evolution of accelerating spacetimes that were somewhat inaccessible or ambiguous until now.  Namely, we showed that the state changes by an order one factor after a Page time has elapsed even in de Sitter, verifying the findings of \cite{1005.1056,1104.0002,1109.1000}. Let us stress that this breakdown is in our standard method of description of the system, as someone who has arranged the initial state in a flat coordinate system would notice the global system becoming incongruously warped on this timescale.  Each observer naturally gauges away these effects locally, and so this does not signify a breakdown in physically observable quantities unless the observer has access to information from many patches (or for long enough times).  This regime is unamenable to our standard perturbative description, but this does not imply that a more appropriate way of organizing the description of the evolution does not exist.  We will return to this point in future work.

\section*{Acknowledgments}
We thank  S.B. Giddings, N. Kaloper, M. Mirbabayi, G.L. Pimentel, and A. Westphal for many interesting discussions.
M.S.S. is supported by the Lundbeck foundation and VILLUM FONDEN grant 13384.
CP3-Origins is partially funded by the Danish National Research Foundation, grant number DNRF90.

\appendix

\section{$\g$ correlators} \label{hcorrelators}

\subsection{Mode function}
We Fourier transform the graviton field as
\begin{eqnarray} \label{hk}
\g_{ij}(x)&=& \int \frac{d^3 k}{(2 \pi)^{3/2}} e^{ik x} \g_{ij}(k) \nn \\
\g_{ij}(k)&=& \sum_{s=s_1,s_2} \epsilon^s(k)_{ij}\g (k) a_k^s + \epsilon^{s}(k)^*_{ij}\g(k)^*a^{s\,\dagger}_{-k}
\end{eqnarray}
and satisfying the canonical commutation relations
\begin{eqnarray}
\left[ a^s_k, a^{s'}_{-k'} \right]= \delta_{s s'} \delta^{(3)}(k+k').
\end{eqnarray}

Gravitons in a flat FLRW space time background satisfy the equation of motion
\begin{eqnarray}
\ddot{\g_k} + 3 H \dot \g_k +k^2 \g_k =0.
\end{eqnarray}
We choose Bunch Davies initial conditions such that the canonically normalized field associated with the graviton behaves as a plane wave at early times. Then, the graviton mode function is given by
\begin{eqnarray} \label{modfunc}
\g_k(\eta)= C_1 (-\eta H) \sqrt{-\eta}  H^{(1)}_\nu(-k \eta),
\end{eqnarray}
where $C_1=\sqrt{\pi/2} e^{i\pi/2(\nu +1/2)}$ and $H^{(1)}$ is the Hankel function of the first kind.

\subsection{Polarization Tensors}

In the main text we deal with correlators of $\{\g,\dg,\g'\}$ where $\g'$ involves $k$ derivatives of $\g$. The correlators not involving $k$ derivatives have the same tensor structure, namely
\begin{eqnarray} \label{C1}
\left<D\,\g_{ij}(k) D\,\g_{mn}(-k)\right> \,= \,P_{ijmn}\langle D\g D\g\rangle
\end{eqnarray}
where $D=\{\partial_t,1\}$ and $P_{ijmn}$ is the projector operator
\begin{eqnarray}
P_{ijmn}(k)= \hat{\delta}_{im} \hat{\delta}_{jn}+\hat{\delta}_{in} \hat{\delta}_{jm}-\hat{\delta}_{ij} \hat{\delta}_{mn}
\end{eqnarray}
where $\hat{\delta}_{ij}= \delta_{ij} - \hat{k}_i \hat{k}_j$.
For correlators involving terms of the form $k_a \partial_{k_b} \gamma_{ij}(k)$ we use the identity \cite{Pimentel:2014nka}
\begin{eqnarray}
k_a \partial_{k_b} (\e_{ij}(k) ) = -\hat k_a \hat k_j \e_{ib}(k) - \hat k_i \hat k_b \e_{aj}(k).
\end{eqnarray}
Using this result and the fact that $\g_{ij}= \g_k e_{ij}(k)$ we get
\begin{eqnarray} \label{kderivative}
k_a \partial_{k_b} \gamma_{ij}(k) = \hat{k}_a \left[\hat{k}_b  \g_k' \e_{ij}- \hat{k}_j \g_{ib} - \hat{k}_i  \g_{bj} \right]. 
\end{eqnarray}

\subsection{Identities} \label{identities}

Here we list several identities used in the main text:
\begin{eqnarray} \label{Id1}
P_{ijmn}P_{ijmn}&=& 8 \\
P_{iamn}P_{ibmc}&=& 2P_{nbac} \\
P_{iamn}P_{ibmn}&=& 4 \hat{\delta}_{ab} \\
\hat{\delta}_{ab} \g^L_{abcd}&=& -\hat{k}_{a}\hat{k}_{b} \g^L_{abcd} \\
P_{najc} \g^L_{ajcn}&=& \hat{k}_a\hat k_j \hat k_c \hat k_n \g^L_{ajcn} \\
P_{ijij}&=&4 \label{id5} \\
\int d\Omega \, \hat{k}_a \hat k_b\hat k_c\hat k_d &=& \frac{4\pi}{15} \left(\delta_{ab}\delta_{cd}+ \delta_{ac}\delta_{bd} + \delta_{ad} \delta_{bc}\right).
\end{eqnarray} 

\section{Overlap technicalities}
\subsection{Resummation}\label{Resummation}
Though we have shown that the state changes by $\mathcal{O}(1)$ on the Page time, from our expression it appears that the overlap becomes large and negative.  This is clearly an artefact of the perturbative expansion we used, and we now show that the overlap tends to 0.  For this we focus on the cubic part of the charge.  In addition to the quadratic contribution we have calculated, there will be an infinite number of higher order contributions due to the exponential series.  Only the ones with an even number of charges will contribute.  It is beyond the scope of the present paper to attempt a full calculation of $\langle Q_3^{2m}\rangle$, but we note that of the various contributions to this quantity, a few of them will be of the form $\langle Q_3^{2}\rangle^m$.  In fact, there will be $(2m)!/2^mm!$ of them.  Then, the exponential series can be resummed as
\beq
\langle e^{iQ}\rangle=e^{i\theta}\left((1+\dots)e^{-\frac{1}{2}\langle Q_3^{2}\rangle}+\dots\right)
\eeq
The ellipsis contains terms that do not involve $Q_3$, terms that involve $Q_3$ as well as other terms in the expansion, and terms only involving $Q_3$ that did not degenerate to products of the quadratic term.  These last terms will act as (an infinite number of) differential operators acting on $e^{-\frac{1}{2}\langle Q_3^{2}\rangle}$.  Though it is a mathematical possibility that these terms conspire to wreck the asymptotic behavior of the exponential, on physical grounds we view this as unlikely.  This indicates that the overlap tends to 0 after the Page time.  Note that while we limited our concern to cubic terms in the charge, higher terms will be suppressed by further powers of $H/M_p$.

\subsection{Contact terms} \label{Contact terms}
In this section we show that contact terms can be resummed into a pure phase, which will not affect the physical evolution of the state.  To see this, let us focus on the first possible terms,
\beq
\langle 0|0'\rangle=1+i\langle Q\rangle-\frac12\langle Q^2\rangle+\dots
\eeq
Naively, the expectation value of $Q$ may look like the appropriate quantity to characterize the effects of long modes.  However, all higher order terms also contain powers of $\langle Q\rangle$, in such a way to render this quantity unphysical.  Concretely, of all the possible Wick contractions of the operators, exactly one will work to yield $\langle Q^n\rangle=\langle Q\rangle^n+\dots$, where the remaining terms contain some level of mixing between spacetime points.  These pure contact terms can then be factored out of the calculation and resummed to yield a pure phase.

This does not quite prove that all terms proportional to $\langle Q\rangle$ assemble into a phase, because there are still `subcontact terms', which involve some correlators that are evaluated at coincident points, and some that are not (see also \cite{1011.0452, 1211.4550} for a discussion on these contact terms).  However, these can be handled by observing that any correlator can be decomposed as
\beq
\langle Q^n\rangle=\langle Q\rangle^n+\sum_{k=0}^{n-2}{n \choose k}\langle Q\rangle^k\langle Q^{n-k}\rangle_\text{NC}
\eeq
where the subscript $\text{NC}$ denotes those correlators which are not purely contact.  From this, we can observe that 
\beq
\frac{\partial}{\partial\langle Q\rangle}\langle Q^n\rangle=n\langle Q^{n-1}\rangle \implies \frac{\partial}{\partial\langle Q\rangle}\langle e^{iQ}\rangle=i\langle e^{iQ}\rangle
\eeq
So that
\beq
\langle e^{iQ}\rangle=F\big(\langle Q\rangle,\langle Q^2\rangle_\text{NC},\langle Q^3\rangle_\text{NC},\dots\big)=e^{i\langle Q\rangle}\tilde{F}\big(\langle Q^2\rangle_\text{NC},\langle Q^3\rangle_\text{NC},\dots\big)
\eeq
The contact terms are factored out as a pure phase, and are physically irrelevant.  The physical effect is dictated by the quadratic terms.  Note that similar arguments cannot be made for the higher order terms, at least in the expression as we have written it, because $\partial\langle Q^m\rangle_\text{NC}/\partial\langle Q^n\rangle_\text{NC}\neq0$.

\subsection{Ergodic Replacement} \label{Ergodicity}

In section \ref{Overlap} we computed the average overlap between vacuum states inside the full inflated region. In this appendix we will show how the same result can be obtained without averaging over the inflated region but, instead, by continuously applying the symmetry transformation on the state, i.e.
\begin{align}
\left| 0(t) \right> = e^{i \int_{a_i H}^{a(t) H} d^3 k_L  \,Q_\xi(k_L)} \left| 0 \right>
\end{align}
where $\xi$ is a diffeomorphism associated with a given long wavelength mode $\g(k_L)$. By integrating over $k_L$, or, equivalently, over diffeomorphisms, we are continuously creating soft modes of momentum $k_L$. This mimics the time evolution of the state where soft modes are continuously exiting the horizon.

The goal is to show that the overlap $\left< 0 | 0(t)\right>$ is equal to the average of the overlap $\left< \left< 0 | 0'\right> \right>$ in the full box.
Similarly to section \ref{Overlap}, we assume that the leading physical contribution comes from the cubic part of the charge, thus, the quantity of interest is
\begin{eqnarray}
\left<\left( \int_{a_i H}^{a(t) H} d^3 k_L  \,Q_\xi (k_L)\right)^2\right>_\nc
\end{eqnarray}
where $Q_\xi(k_L)$ is given in \ref{CubicCharge}.
The computation follows closely the steps in section \ref{Overlap}. After Fourier transforming all fields we get \small
\begin{eqnarray} 
\left<\int_{a_i H}^{a(t) H} \,Q_\xi^2\right>_\nc= \left(\frac{a^3 M_p^2 }{8}  \right)^2 \int_{a_i H}^{a(t)H} d^3 k_L d^3 q_L \int_{V^{-1}} d^3 k d^3 q    \Big< \dot{\g_{ij}} (-k)  D_L(k) \g_{ij} (k)\dot{\g_{mn}} (-q)  D_L(q) \g_{mn} (q)  \Big>_\text{NC}.
\end{eqnarray} \normalsize
Although we integrate over long modes the separation of scales remains, i.e, $k_L,q_L \ll k,q$. Therefore, on-shell the long modes contract with each other and we are left with
\begin{eqnarray}
\left<\int_{a_i H}^{a(t) H} \,Q_\xi^2\right>_\nc &=& \left(\frac{a^3 M_p^2 }{8}  \right)^2 \int_{a_i H}^{a(t)H} d^3 k_L \int_{V^{-1}} d^3 k \nn \\ && \mkern-88mu\left[ \left< \dot{\g_{ij}} (-k-k_L)  \dot{\g_{mn}} (k+k_L )\right>'  \left<  \g_{ab}(k_L) \g_{de}(-k_L) \right> \left< k_b \partial_{k^a} \g_{ij} (k) (-q_d) \partial_{-q^e} \g_{mn} (-q) \right>' + \right. \nn \\ &&  \mkern-88mu+ \left. \left<\g_{ab}(k_L) \g_{de}(-k_L) \right>  \left<  \dot{\g_{ij}} (-k-k_L)   (-k_d) \partial_{-k^e} \g_{mn} (-k)\right>' \left< k_b \partial_{k^a} \g_{ij} (k)  \dot{\g_{mn}} (k+k_L)\right>'  \right]\nn
\end{eqnarray}
where the prime in the correlators means that the delta function was extracted.
In the limit where $k_L \ll k$ we recover \ref{CubicCharge_preint} but now integrated over the soft modes $k_L$
\begin{eqnarray}
\left<\int_{a_i H}^{a(t) H} \,Q_\xi^2\right>_\nc = 16 V \left(\frac{a^3 M_p^2 }{8} \right)^2 \int_{a_i H}^{a(t)H} d^3 k_L \left< \g^L_{ab} \g^L_{cd} \right> \int_{V^{-1}} d^3 k \,  \hat{k}_a\hat k_b\hat k_c\hat k_d    |\dg|^2 |\g_k|^2  .
\end{eqnarray}
Given that $|\g_{k_L}|^2=H^2(1+(k_L \eta)^2)/(M_p^2 k^3)$ and that $k_L \eta \ll 1$ eq. (\ref{Double Average of Q}) is recovered.

\section{Renormalization of the charge} \label{Renormalization}

\subsection{Preliminaries}
In computing the overlap between vacua which differ by a charge transformation we encountered divergences in the momentum integral. One divergence comes from the IR cutoff of the charge, as discussed in \ref{V div}. The divergence appears when we allow the lower limit of integration, the radius of the sphere, to be free (we parameterize it by b: $r=b/(aH)$) and take $b \rightarrow +\infty$. The second divergence comes from the upper limit in the momentum integration and naturally corresponds to the UV cutoff. This divergence is also expected, as it is associated with contact terms and so local physics.

In order to renormalize this charge we will require the following criteria to be satisfied:
\begin{itemize}
	\item The same procedure should regulate all the divergences;
	\item The final expression $\langle Q^2 \rangle$ should be positive, since otherwise the overlap would tend towards infinity at late times, rather than zero;
\end{itemize}

The strategy we use to regulate the charge is to add gravitational counterterms in the boundary up to dimension 4. Namely, we will add  to the Lagrangian the following counterterms
\begin{eqnarray} \label{ActionCounterTerms}
S &= & M_p^2 \left[ \int d^4 x \sqrt{-g^{(4)}} \frac{R}{2} + \int_{\cI} d^3 x \sqrt{-g^{(3)}} \left(- K[g] \right. \right. \nn \\ && \underbrace{ \left. \left. -2 c_1 H + \frac{c_2}{2} R[g] +  c_3 R[g]^2+ c_4 R[g]_{ij} R[g]^{ij}   \right) \right]}_\text{Counterterms} 
\end{eqnarray}
where $g$ denotes the induced 3-metric $g_{ij}$ and $c_i$ are the coefficients of the counterterms (with the appropriate dimensions). Note that we did not include the Riemann tensor square because in 3 dimensions it can be written in terms of $R$ and $R_{ij}$. The Brown-York tensor can then be obtained by varying the action with respect to the induced metric $g_{ij}$ on the boundary leading to \cite{1302.2151}
\begin{eqnarray}
T_{ij}^{BY}=\frac{2 }{\sqrt{-g}} \frac{\delta S}{\delta g^{ij}}&=& M_p^2 \left[K_{ij} - K g_{ij} - 2 c_1 H \g_{ij}- c_2 G_{ij} -  \left(c_3 R^2+ c_4 R_{ab} R^{ab}\right)  g_{ij} + 2 c_4 \Delta R_{ij} \right. \nn \\
&& \left.   +4\left(  c_3 R R_{ij} - c_4 R^{de} R_{diej}  \right) - 2\left((2c_3 + c_4) \nabla_i \nabla_j R - (2c_3+c_4/2) g_{ij} \Delta R   \right)  \right]\nn
\end{eqnarray}
where $R, R_{ij}$ and $R_{ijmn}$ are constructed from the induced metric $g_{ij}$, $\nabla_i$ is the covariant derivative associated with $g_{ij}$ and $G_{ij}$ is the Einstein tensor.
This approach is inspired by the regularization of the Brown-York tensor done both in AdS \cite{hep-th/9902121, hep-th/0002230} and dS \cite{hep-th/0110108, hep-th/0110087, 1009.4730}.

We renormalize $\left< Q^2 \right>$ order by order in the counterterms. That is, while the terms linear in $R$ (partially) cancel the divergences of the bare charge squared, we also need to consider $R^2, R_{ij}R^{ij}$ terms on the boundary to completely renormalize our charge.  Note that even after this, quadratic terms in the highest order counterterms will induce divergences that must then be cancelled by cubic curvature terms, and that this process will not terminate, due to the nonrenormalizability of gravitational theories.
Dimensionally, and following this procedure, we would only need the first two counterterms to regulate both the UV and IR divergences (they come together in most of the terms). However, it turns out that the leading UV divergence in the $c_2$ counterterm cancels out. Therefore, in order to cancel the divergences in the original charge the coefficient $c_2$ would have to be $ \cO (\Lambda^2/H)$ which would drive the theory ill-behaved. On the other hand, while in \cite{1009.4730} the IR divergences were canceled by $c_2$, it is natural to expect that $UV$ divergences should be canceled by different operators which, in the case of gravity, typically have to be higher dimensional. For that reason we consider dimension four operators.

Before proceeding let us first note that the first counterterm has already been used to show the equivalence between the Noether and the Brown-York charge. Namely, when performing the canonical transformation between the 3-metric and its conformally related metric, one extra term would appear in the charge coming from time derivatives of the scale factor. However, this term can be cancelled by subtracting a cosmological constant on the boundary, namely, by fixing $c_1=1$.

In what follows we will need expressions for the Ricci scalar and Ricci tensor up to second order in $\g$ and for the Riemann tensor up to first order in $\gamma$. They are given by
\begin{eqnarray}
R_{ijlm}&=& R_{ijlm}^{(1)}  +\cO (\g^2)\\
R_{ij}&=& R^{(1)}_{ij} + R^{(2)}_{ij} +\cO(\g^3)  \\
R &=& R^{(2)} +\cO(\g^3)
\end{eqnarray}  
where the upper index denotes the order in $\g$.
More explicitly,
\begin{eqnarray} \label{Riemann}
R_{ijlm}^{(1)}&=& \frac 12 \left[\g_{im,jl} + \g_{jl,im} - \g_{jm,il} - \g_{il,jm} \right] \nn\\ \label{Ric1}
R^{(1)}_{ij}&=& -\frac{\g_{ij,kk}}{2} \nn\\ \label{Ric2}
R^{(2)}_{ij}&=& -\frac{1}{4} \left[ \gamma_{ak} \left(  \g_{(ia,j)k}  - 2\g_{ij,ak} \right) -  \g_{(ib,k} \g_{bk,j)}   + \g_{(ib} \g_{bj),kk} + 2\gamma_{ik,m} \gamma_{jm,k} + \gamma_{mk,i} \gamma_{mk,j} \right]\nn\\ \label{R2}
R^{(2)}&=&  -\frac{1}{4a^2} \gamma_{mk,i} \gamma_{mk,i} 
\end{eqnarray}
where $(,)$ denotes symmetrization in the outermost indices, i.e. $O_{(iaj)}=O_{iaj}+O_{jai}$.  We will only ever symmetrize over two indices, so the notation is unambiguous.

We are ultimately interested in the cubic charge. The contribution of the counterterms to the cubic charge is of the form
\begin{eqnarray}
Q_3= \frac{a^3}{2} \int d^3x \,T^{ij} \delta g_{ij}. 
\end{eqnarray}
Note that here we are still using the 3-metric $g_{ij}$ and not its conformally related metric.

\subsection{$R^1$ terms}
We are now in position to start the bulk of the computation. As we explained before, the first counterterm cancels the cosmological constant terms and does not add any new contributions. Therefore, we can focus on remaining counterterms. We start by computing the charge associated with $c_2 R$. Its contribution to the cubic charge, which we denote by $Q_3|_\text{CT2}$, is then given by
\begin{eqnarray}\label{CT_charge}
Q_3|_\text{CT2}= -\frac{c_2 M_p^2}{2a} \int d^3 x \left[-\gamma_{(ia} R^{(1)}_{aj)} \delta g^{(1)}_{ij} + R_{ij}^{(1)} \delta g^{(2)}_{ij} +\left( R_{ij}^{(2)} -\frac{a^2}{2} R^{(2)} \delta_{ij} \right) \delta g^{(1)}_{ij} \right]
\end{eqnarray}
where again the upper index denotes the order in $\g$. As we can see from \ref{deltagij} the variation of the metric has linear terms in $\xi$ (which are proportional to $\g_L$) and terms in $\xi, \g_{ij}$ which will be the quadratic pieces\footnote{In full generality there will also be terms in $\xi$ quadratic in the long mode. However, those terms will force all fields in the charge to be long (at cubic order) and so for our finite volume they do not contribute. Therefore, we here consider only the leading term, $\xi= \g^L_{ij} x_j/2$.}
\begin{eqnarray}
\delta g_{ij}^{(1)}&=& a^2\partial_{(i} \xi_{j)}\\
\delta g_{ij}^{(2)}&=& a^2\xi\cdot\partial\gamma_{ij}.
\end{eqnarray}
Using \ref{deltagij} and \ref{R2} yields
\begin{eqnarray} \label{CT2Q1}
Q_3|_\text{CT2}= -\frac{a c_2 M_p^2}{2} \int d^3 x  \left[  \gamma_{ia} \g_{aj,kk}  \partial_{(i} \xi_{j)} - \frac 12 \g_{ij,kk} \xi\cdot\partial\gamma_{ij}  + \left( R_{ij}^{(2)} +  \frac{1}{8} \gamma_{mk,b} \gamma_{mk,b}  \delta_{ij} \right) \partial_{(i} \xi_{j)} \right]\nn
\end{eqnarray}
Note that due to the traceless condition, and for $\xi= \g^L_{ij} x_j/2$, some terms disappear leaving 
\begin{eqnarray} \label{CT2_Q2}
Q_3|_\text{CT2}= \frac{a c_2 M_p^2}{2} \int d^3 x \left[   \gamma_{ia} \g_{aj,kk}  \g^L_{ij} - \frac 14 \g_{ij,kk} \g^L_{ab} x_b \partial_a \gamma_{ij} + R_{ij}^{(2)} \g^L_{ij}  \right].
\end{eqnarray}
The next step is to Fourier transform the charge. Given the complexity of the expression for $R^{(2)}_{ij}$ it is recommendable to take closer look at the several terms. Namely, in the limit where the moment of the long mode is taken to zero the Ricci tensor has schematically the form $ k^2  \gamma(k) \gamma(-k)$. Therefore, each time a derivative is contracted with $\g$ it will give zero by the transverse condition. The only surviving terms of \ref{Ric2} are the last and third to last. After Fourier transforming and using the simplification just mentioned we arrive at 
\begin{eqnarray} \label{CT2_Q3}
Q_3|_\text{CT2}= \frac{a c_2 M_p^2}{2} \int d^3 k \, k^2 && \left[  - \gamma_{ia} (k) \g_{aj} (-k)  \g^L_{ij} - 
 \frac 14 \g_{ij}'(-k)  \g^L_{ab} \hat{k}_b \hat{k}_a \gamma_{ij}(k) +   \right.\nn \\ && + \left.  \frac 12 \left(\gamma_{ia}(k) \g_{aj} (-k) -\frac 12 \hat{k}_i \hat{k}_j \g_{bm} (k) \g_{bm} (-k) \right) \g^L_{ij}  \right],
\end{eqnarray}
which simplifies to
\begin{eqnarray} \label{CubicChargeCT}
Q_3|_\text{CT2}= -\frac{a c_2 M_p^2}{8} \int d^3 k \, k^2 \left[   2 \gamma_{ia} (k) \g_{aj} (-k)  \g^L_{ij} + 
\g^L_{ab} \hat{k}_b \hat{k}_a \g_{ij}(-k)\left(  \g_{ij}'(k) + \g_{ij}(k)\right) \right].
\end{eqnarray}

\subsection{$R^2$ Terms}
We now pass to the charge associated with the higher dimensional terms $R^2$ and $R_{ij}^2$. Lets look separately at its contributions to the Brown-York tensor 
\begin{eqnarray}
T_{ij}^{BY}|_{CT}&=& 2 M_p^2 \left[ \frac 12 \left(c_3 R^2+ c_4 R_{ab} R^{ab}\right)  g_{ij}  -2\left(  c_3 R R_{ij} + c_4 R^{de} R_{diej}  \right) \right. \nonumber\\ && \left. + (2c_3 + c_4) \nabla_i \nabla_j R - \left(2c_3+\frac{c_4}{2}\right) g_{ij} \Delta R - c_4 \Delta R_{ij}  \right].
\end{eqnarray}
In terms of $\gamma$: $R_{abcd}, R_{ij} \simeq \cO (\g)$ and $R \simeq \cO (\g^2)$. Therefore, there are no zero order terms and only the last term contributes to first order in $\g$ . Terms in $R^2$ and $R\, R_{ij}$ are $\cO (\g^3)$ or smaller so we neglect them. Thus, up to second order in $\g$ we have
\begin{eqnarray} \label{BY_CT3}
T_{ij}^{BY}|_{CT}= \frac{2 M_p^2}{a^2} &&\left[\frac 12 c_4 R_{ab}^{(1)} R^{ab\,(1)} \delta_{ij}  -2 c_4 R^{de\,(1)} R_{diej}^{(1)}  + a^2 (2c_3 + c_4)  R_{,ij}^{(2)} - \right. \nn \\ &&  \left. - a^2 \left(2c_3+\frac{c_4}{2}\right) \delta_{ij}  R^{(2)}_{,kk} - c_4 \left(R_{ij,kk}^{(1)} + R_{ij,kk}^{(2)} -\g^{kb} R_{ij,kb}^{(1)}  \right)   \right]
\end{eqnarray}
where in the last equation indices are lowered and raised with the identity matrix.
Fortunately, using the knowledge from previous computations we know that terms of the form $\delta_{ij} \delta g_{ij}^{(1)}$ vanish. Therefore, the first and fourth term of the previous expression do not contribute to the charge. Moreover, most of the terms will have the exact same form as those in \ref{CT_charge} times $k^2$ which allows us to simply copy some of the previous expressions.
Let us first rewrite the new contribution to the charge from \ref{BY_CT3} in the form of \ref{CT_charge}
\begin{eqnarray} \label{CT_Q}
Q_3|_\text{CT3}&=& \frac{1}{2 a} \int d^3x \left[ - \g_{(ib} T^{(1)}_{bj)}  \delta g_{ij}^{(1)} + T_{ij}^{(2)} \delta g_{ij}^{(1)} + T_{ij}^{(1)}  \delta g_{ij}^{(2)} \right] \nn \\
&=&  \frac{M_p^2}{a^3} \int d^3x \left[ -c_4 a^2 \left(- \g_{(ib} R^{(1)}_{bj),kk} \p_{(i} \xi_{j)}+ R^{(1)}_{ij,kk} \xi_a \partial^a \g_{ij} \right)  + \tilde T_{ij}^{(2)} \delta g_{ij}^{(1)}  \right]
\end{eqnarray}
The first term ultimately vanishes from the quantity we calculate, because of the tracelessness of $\g^L$.  The terms coming from $\tilde T_{ij}^{(2)}$ are
\begin{eqnarray}
\tilde T_{ij}^{(2)} \delta g_{ij}^{(1)}= - \left[c_4 \left(R_{ij,kk}^{(2)}- \g_{kb} R_{ij,kb}^{(1)} \right)- a^2(2c_3 +c_4) R_{,ij}^{(2)} + 2 c_4 R_{de}^{(1)}R_{diej}^{(1)} \right] \delta g_{ij}^{(1)}
\end{eqnarray}
The first and third terms vanish because they are total derivatives, the second vanishes due to transversality, and so we are only left with the term involving the Riemann tensor. A further simplification occurs due to transverse conditions in Fourier space such that only the last term in the expression for the Riemann tensor in \ref{Riemann} survives and, using \ref{Ric1}, we are left with
\begin{eqnarray}
Q_3|_\text{CT3}=\frac{M_p^2c_4}{a} \int d^3x\left(-\frac12 \Delta^2\g_{de}D_L\g_{de} +\frac12 \Delta\g_{de} \gamma_{de,ij} \g^L_{ij}\right).
\end{eqnarray}
After Fourier transforming, this gives
\begin{eqnarray} \label{Q_CT3}
Q_3|_\text{CT3}=\frac{3}{4} c_4 a^3 M_p^2  \int d^3k \left(\frac{k}{a}\right)^4   \hat{k}_i \hat{k}_j \g_{de} \g_{de} \g^L_{ij}. 
\end{eqnarray}
Now that we have an explicit and simplified expression for $Q_3|_\text{CT2,CT3}$ we can proceed to renormalize $\left< Q^2 \right>$. The counter terms have a similar form to the original charge in Fourier space (c.f. \ref{CubicCharge_kspace}) apart from an overall $(k/a)^2, (k/a)^4$ dependence. In particular the second term in \ref{CubicChargeCT} and \ref{Q_CT3} have exactly the same angular dependence.

\subsection{Result}
As we stated before, we renormalize $\left< Q^2 \right>$ order by order in the counter terms. This gives
\begin{eqnarray}
\left<Q^2\right>_\text{Ren} = \left<Q^2 +\{Q, Q_\text{CT2}\} +Q_\text{CT2}^2+  \{Q, Q_\text{CT3}\} \right>. 
\end{eqnarray}
Terms of the form  $\int d^3x \hat{k}_a \hat{k}_b \g^L_{ab} \propto \g^L_{aa}$ vanish. Thus, using \ref{CubicCharge_preint} and \ref{CubicChargeCT} we simply get
\begin{eqnarray}
\left<Q^2\right>_\text{Ren} = 8\pi \frac{16 }{15}  V \left(\frac{a^3 M_p^2 }{8} \right)^2 \g^L_{ab} \g^L_{ab}  \int_{\frac{aH}{b}}^\Lambda dk k^2  \left[ |\dg|^2 \left| \g'_k\right|^2- 2 c_2 \left(\frac{k}{a}\right)^2 \text{Re}\left[  \dg \g_k^* \left| \g'_k\right|^2 + \g_k^2 \dot{\g_k}^* \g_k'^* \right] \right. \nn \\  \left. + c_2^2 \left(\frac{k}{a}\right)^4\left|\g_k \right|^2 \left( \left| \g_k' \right|^2+2\left| \g_k \right|^2   + 2 \text{Re}\left[\g_k \g_k'^*\right] \right) - 12c_4 \left(\frac{k}{a}\right)^4   \text{Re}  \left[ (\g_k^*)^2 \dot{\g}_k \g_k'  \right]\right]\nn
\end{eqnarray}
The previous expression has divergences both in $b$ and in $\Lambda$ when taking each one to infinity. We regularize this expression by choosing $c_2$ such that there are no IR divergences, i.e., the strongest b-dependence in the $k$ integral is $b^{-3}$. This procedure also regulates the UV divergences. In fact, regulating the $b$ divergences is a stronger condition, as there are terms diverging in $b$ but not in $\Lambda$.  The scheme ambiguity shows up here by our choice of enforcing that the divergent terms vanish.  If instead we only insist that they yield a finite value, this would simply shift the overall coefficient of $\left<Q^2\right>_\text{Ren} $, while leaving the choices of counter terms invariant.  This procedure gives,to leading order in the cutoff , $c_2= \pm \sqrt{12 H c_4} $ (it agrees with \cite{1009.4730} if $c_4=1/(\sqrt{12}H^3)$ and the final result is
\begin{eqnarray} \label{QRen}
\left<Q^2\right>_\text{Ren} = 8\pi \frac{16 }{15}  V_H \left(\frac{a^3 M_p^2 }{8} \right)^2 \g^L_{ab}\g^L_{ab} \left(\frac{H^3}{4 a^3 M_p^4}\right)~,
\end{eqnarray}
where $V_H\equiv 1/(aH)^3$.

Note that the addition of these counter terms to the charge does not affect previous results, like the consistency conditions, as the effect of the counter term is suppressed by $k/(aH)$.

\bibliographystyle{JHEP}
\bibliography{AsympSym}

\end{document}